\documentclass[useAMS,usenatbib]{mn2e}

\usepackage{graphicx,mathptm}
\usepackage{times}%
\usepackage{natbib}

\newcommand{\mincir}{\raise
  -2.truept\hbox{\rlap{\hbox{$\sim$}}\raise5.truept \hbox{$<$}\ }}
\newcommand{\magcir}{\raise
  -2.truept\hbox{\rlap{\hbox{$\sim$}}\raise5.truept \hbox{$>$}\ }}
\newcommand{\siml}{\raise
  -2.truept\hbox{\rlap{\hbox{$\sim$}}\raise5.truept \hbox{$<$}\ }}
\newcommand{\simg}{\raise
  -2.truept\hbox{\rlap{\hbox{$\sim$}}\raise5.truept \hbox{$>$}\ }}

\title[Simulating the physical properties of dark matter and 
gas inside the cosmic web]
{Simulating the physical properties of dark matter and 
gas inside the cosmic web}
\author[K. Dolag et al.]
{K. Dolag$^{1}$, M.  Meneghetti$^{2,3}$, L. Moscardini$^{4}$,
E. Rasia$^{5,4}$ and A.  Bonaldi$^5$\\
$^{1}$ Max-Planck Institut fuer Astrophysik,
Karl-Schwarzschild Strasse 1, D-85748 Garching, Germany
(kdolag@mpa-garching.mpg.de)\\
$^{2}$ Zentrum fuer Astronomie, ITA, Universitaet Heidelberg,
Albert-Ueberle-Strasse 2, D-69120 Heidelberg, Germany
(meneghetti@ita.uni-heidelberg.de)\\
$^{3}$ INAF, Osservatorio Astronomico di Bologna,
via Ranzani 1, I-40127 Bologna, Italy \\
$^{4}$ Dipartimento di Astronomia, Universit\`a di Bologna,
via Ranzani 1, I-40127 Bologna, Italy
(lauro.moscardini@unibo.it)\\
$^{5}$ Dipartimento di Astronomia, Universit\`a di
Padova, vicolo dell'Osservatorio 2, I-35122 Padova, Italy
(rasia,bonaldi@pd.astro.it) \\
}
\begin{document}

\date{Accepted ???. Received ???; in original form November 2005}

\pagerange{\pageref{firstpage}--\pageref{lastpage}} \pubyear{2006}

\maketitle

\label{firstpage}


\begin{abstract}
Using the results of a high-resolution, cosmological hydrodynamical
re-simulation of a supercluster-like region we investigate the
physical properties of the gas located along the filaments and bridges
which constitute the so-called cosmic web. First we analyze the main
characteristics of the density, temperature and velocity fields, which
have quite different distributions, reflecting the complex dynamics of
the structure formation process. Then we quantify the signals which
originate from the matter in the filaments by considering different
observables. Inside the cosmic web, we find that the halo density is
about 10-14 times larger than cosmic mean; the
bremsstrahlung X-ray surface brightness reaches at most $10^{-16}$ erg
s$^{-1}$ cm$^{-2}$ arcmin$^{-2}$; the Compton-$y$ parameter due to the
thermal Sunyaev-Zel'dovich effect is about $10^{-6}$; the reduced
shear produced by the weak lensing effect is $\sim 0.01-0.02$.  These
results confirm the difficulty of an observational detection of the
cosmic web. Finally we find that projection effects of the filamentary
network can affect the estimates of the properties of single clusters,
increasing their X-ray luminosity by less than 10 per cent and their
central Compton-$y$ parameter by up to 30 per cent.
\end{abstract}



\section{Introduction} \label{sect:introduction}

In the standard picture of structure formation, the large scale
structure observed in the spatial distribution of galaxies has evolved
via gravitational instability from small perturbations present in the
primordial cosmic density field.  In particular, the later non-linear
stages of this evolutionary process are thought to be responsible for
the complex pattern of large voids surrounded by galaxies, clusters
and superclusters exhibited by the observed galaxy distribution in
modern catalogues, like the 2dF and SDSS surveys \citep[see,
e.g.,][]{peacock01,tegmark04}.

This gravitational instability scenario has been largely confirmed by
the perturbation theory in the weakly non-linear regime and by the
results of numerical N-body simulations. In the cold dark matter (CDM)
cosmological model, cosmic structures form by the accretion of smaller
sub-units in a hierarchical bottom-up fashion. In this framework,
galaxy clusters are of fundamental importance, since they are the
largest bound systems in the universe and those most recently
formed. Moreover, their abundances and spatial distribution retain the
imprints of the underlying cosmological model; it is therefore
possible to constrain the fundamental parameters of the model by
studying the cluster properties and their redshift evolution
\citep[see][ and references therein]{rosati02a,voit2005}.

Still, the theoretical picture emerging from numerical simulations is
even more intriguing. The largest dark matter haloes, which are
expected to host the richest galaxy clusters, seem to be
preferentially located at the intersections of one-dimensional,
filamentary structures having a much smaller density contrast of the
order of unity.  From a dynamical point of view, this network of
filaments (the so-called cosmic web) represents the preferential
directions along which the matter streams to build the massive and
dense structures located at the nodes. This picture, first emerged
from N-body experiments, was lately understood in a theoretical
framework \citep[see][]{bond96}, in which the cosmic web corresponds
to the density enhancements present in the primordial field that are
then sharpened by the following non-linear evolution
\citep[see also][]{doroshkevich1982}.  A firm prediction of this
model is that filaments represent bridges between denser spots
identified with galaxy clusters.

The observational confirmation of this general picture resulted to be
challenging. Some evidence arose from the presence of elongated
structures such as the Great Wall \citep{geller1989}, in the galaxy
spatial distribution, and from the tendency of neighboring clusters to
be aligned \citep[see, e.g.][]{plionis2003}.  More recent attempts at
detecting filaments which make use of specific statistical analyses of
modern galaxy surveys gave new support for the existence of the cosmic
web \citep[see, e.g.,][]
{bardelli00,ebeling04,pimbblet04a,pimbblet04b,bharadwaj2004,pimbblet2005,
porter2005,pandey2005}.  Further evidence has been found by studying the
ultraviolet absorption line properties of active galactic nuclei
projected behind superclusters of galaxies \citep[][]{bregman2004}.

Hydrodynamical simulations, including the treatment of the gas
component, suggest that regions of moderate overdensity like filaments
host a significant fraction of gas with a temperature between
$10^5$ and $10^7$ K, the so-called warm-hot intergalactic medium
(WHIM) \citep[see, e.g.][]{cen99,pierre2000,croft01,dave01,phillips01}.
This gas may reveal its presence through its bremsstrahlung emission
in the soft X-ray band. However, the expected signal appears too small
to be clearly detected by the present generation of satellites: in the
literature there are only a few claims of possible detections of WHIM
emission
\citep{scharf00,zappacosta02,markevitch03,kaastra03,finoguenov03}.

Much effort has been put into investigating the overall contribution
of diffuse gas to the soft X-ray background \citep[see,
e.g.][]{croft2001,roncarelli2006}. A major issue is to distinguish it
from that of the two dominant sources, i.e. the active galactic nuclei
and the hot gas in our own Galaxy.  Thanks to deep observations with
the Chandra satellite
\citep{brandt01,rosati02b,bauer2004,worsley2005}, it has become clear
that at least 90 per cent of the background in the (0.5-2 keV) band is
contributed by discrete sources, thus implying an upper limit to the
WHIM contribution. This result can be used to constrain the thermal
properties of baryons and their cosmic history, providing important
information on the evolution of galaxy clusters
\citep[see, e.g.,][]{voit01,bryan01,xue03}.

An alternative way to investigate the diffuse baryons in filaments is
the Sunyaev-Zel'dovich (SZ) effect  \citep[see, e.g., the
simulated maps obtained by][]{dasilva2001,springel2001,white2002}.
The SZ and bremsstrahlung signals have a different dependence on the
gas density and thus can be used to investigate the gas properties in
two different but complementary environments. An additional probe of
the WHIM at intermediate angular scales and redshifts is enabled by
the study of the cross-correlation between soft X-ray and SZ maps,
like those which will be obtained by planned missions in the 
future \citep{cheng04}.

Finally, gravitational lensing represents a unique tool for
constraining the mass distribution of the cosmic structures.  On
galaxy cluster scales, strong gravitational lensing is used for
measuring the mass in the inner regions of the lenses, where highly
distorted images of background galaxies appear in form of arcs. In the
outer regions of clusters, where the surface density is smaller, the
gravitational lensing effects manifest themselves as small changes of
the source shapes and orientations, which can be detected only by
averaging over a large number of galaxies in a finite region of the
sky. Due to the intrinsic ellipticities of the background galaxies,
the weak lensing approach is unfortunately limited by noise but still
represents a powerful method for reconstructing the lens mass
distribution. Compared to other techniques, which make use of the
light for tracing the mass, the largest advantage comes from the fact
that it is model-independent and that no strong assumptions are
required when converting the lensing signal into a mass measurement.
Although this method has worked successfully in reconstructing the
mass distribution around many galaxy clusters, it has failed yet to
detect shallow surface density structures like the cosmic web. It is
expected that the typical surface mass density along the filament is
too small for producing a weak lensing signal detectable with the
current instruments \citep{jain00}. However, as suggested by
\cite{dietrich2005}, the effect might be measurable if the clusters
connected by the filament are nearby.  A few candidates of filaments
have been found through weak lensing so far, but without strong
constraints. There is some evidence of a bridge connecting member
clusters in the Abell 901/902 \citep{gray02} and MS0302+17
\citep{kaiser98,gavazzi04} superclusters. \cite{dietrich2005}
recently claimed the detection (at the $2\sigma$ level) of a
filamentary structure between the two massive clusters A222 and
A223. Their result is also supported by the good correlation with the
optical and X-ray data. Recently, \cite{porter2005} identify
two major filaments in the Pisces-Cetus supercluster at the edges of
the 2dF Galaxy Redshift Survey and the Sloan Digital Sky survey.

Although, many observations of filamentary structures have been
attempted using a variety of techniques, as described earlier, it is
still unclear how such filaments may appear in many of their
observables. This makes it difficult to compare different
observations, mainly because they may be efficient in probing
different constituents of the cosmic web.  The main goal of this paper
is to cover this gap to give a detailed overview of how filamentary
structures look like from as many points of view as it is possible. We
focus our attention on a restricted region of a cosmological
simulation, which contains a supercluster-like structure extending
over several Mpc. We perform a new high-resolution hydrodynamical
simulation of this sub-volume, which represents one of the largest
overdensities found in the original cosmological box. The region
contains four galaxy clusters with a mass larger than about $10^{15}
M_{\odot}$, connected by filaments and bridges. We discuss the
properties of these structures with respect to many observables.

The plan of the paper is as follows. In Section~\ref{sect:simulation}
we describe the general characteristics of the simulation used in the
following analysis.  In Section~\ref{sect:cosmicweb} we introduce the
concept of cosmic web and bridges, discussing their morphological
appearance.  Section~\ref{sect:gasproperties} is devoted to the
discussion of the main properties of the gas inside the bridges: in
more detail we analyze its density, temperature and velocity field.
In Section~\ref{sect:observations} we discuss the observational
implications of the previous results: in particular we study the
density and spatial distribution of haloes, the characteristics of the
bremsstrahlung emission in the X-ray band, the Compton-$y$ parameter
produced by the thermal SZ effect, and the weak gravitational lensing
signal.  Finally, we summarize our results and draw our main
conclusions in Section~\ref{sect:conclusions}.


\section{The numerical simulation} \label{sect:simulation}

The numerical simulation discussed in this paper follows the formation
and evolution of a supercluster-like structure, which at the present
epoch contains 27 haloes with virial mass $M_{\rm vir}$ larger than
$0.7\times 10^{14} M_\odot$. Four of them are massive galaxy clusters
with $M_{\rm vir}\magcir 10^{15} M_\odot$, connected by bridges and
sheets of gas and dark matter.  This overdense region was suitably
chosen in the final output of a parent cosmological N-body simulation
\citep{yoshida01,jenkins01}, which follows the evolution of the dark 
matter (DM) component only in a box with size $479\,h^{-1}$Mpc comoving; the
assumed background cosmology is the standard `concordance' cosmogony,
i.e. a flat $\Lambda$CDM model with a present matter density parameter
$\Omega_{\rm 0m}=0.3$, a present baryon density parameter $\Omega_{\rm
0b}=0.04$, a Hubble parameter $h=0.7$, and a power spectrum
normalization corresponding to $\sigma_8=0.9$.

By adopting the `Zoomed Initial Conditions' (ZIC) technique
\citep{tormen97}, the corresponding Lagrangian region in the initial
conditions was resampled to higher resolution by appropriately adding
small-scale power and increasing the number of particles.  This allows
the mass and force resolution to be increased almost at will.  In
order to reproduce the global tidal field, outside the high-resolution
(HR) region the primordial fluctuations are still sampled, but with a
reduced number of particles having larger mass. The exact shape of the
HR region has to be accurately selected to avoid any contaminating
effect from these low-resolution (LR) particles.  This was done using
an iterative process as follows.  Starting from a first guess of the
HR region, we run a LR DM-only re-simulation. Analyzing its final
output, we individuate in the initial conditions the Lagrangian region
containing all the particles that are at a distance smaller than 5
virial radii ($R_{\rm vir}$) from at least one of the 27 main haloes.
Applying ZIC, we generate new HR initial conditions and run one more
DM-only re-simulation.  The procedure is iteratively repeated until we
find that none of the LR particles enter the HR region, which would be
possible because of the introduction of small-scale modes. Because of
the complex geometry of the supercluster region considered, in our
case this procedure required more than 20 low-resolution DM-only
re-simulations.  The final HR region has an irregular shape which
can be contained into a box of roughly $50\times50\times70$
Mpc$^3$. All 27 haloes are free from contaminating boundary effects
up to at least 5 times $R_{\rm vir}$.  These new initial conditions
were then adapted for our hydrodynamical run, performed using
{\small GADGET-2} \citep{springel2005}, a new version of the parallel
TreeSPH code {\small GADGET} \citep{springel01}, which adopts an
entropy-conserving formulation of SPH \citep{springel02}.  Gas
particles have been introduced only in the HR region by splitting each
original particle into a gas and a DM particle with masses $m_{\rm
DM}=1.61\times 10^9\,M_\odot$ and $m_{\rm gas}=2.43\times
10^8\,M_\odot$, respectively.  The Plummer--equivalent gravitational
softening was set to $\epsilon_{\rm Pl} \approx 7$ kpc comoving from
$z=2$ to $z=0$, while it was taken to be fixed in physical units at
higher redshifts.  Our re-simulation, which includes non-radiative
hydrodynamics only, was performed in parallel by using approximately
13,000 hours on 16 CPUs on an IBM-SP4 located at the CINECA
Supercomputing Centre in Bologna, Italy.

Finally, let us to remark that the most important difficulty in
creating the suitable initial conditions and in running the simulation
has been the fact that the structure we want to study, i.e. the cosmic
web, is quite extended and has very low density. Consequently a HR
simulation needs to follow a large volume with a very large number of
particles: in the HR region we used approximately $1.4 \times 10^7$
gas particles and $1.7 \times 10^7$ DM particles, plus $1.4
\times 10^6$ LR particles in the external region.


\section{The cosmic web}\label{sect:cosmicweb}

\begin{figure*}
\begin{center}
\includegraphics[width=0.865\textwidth]{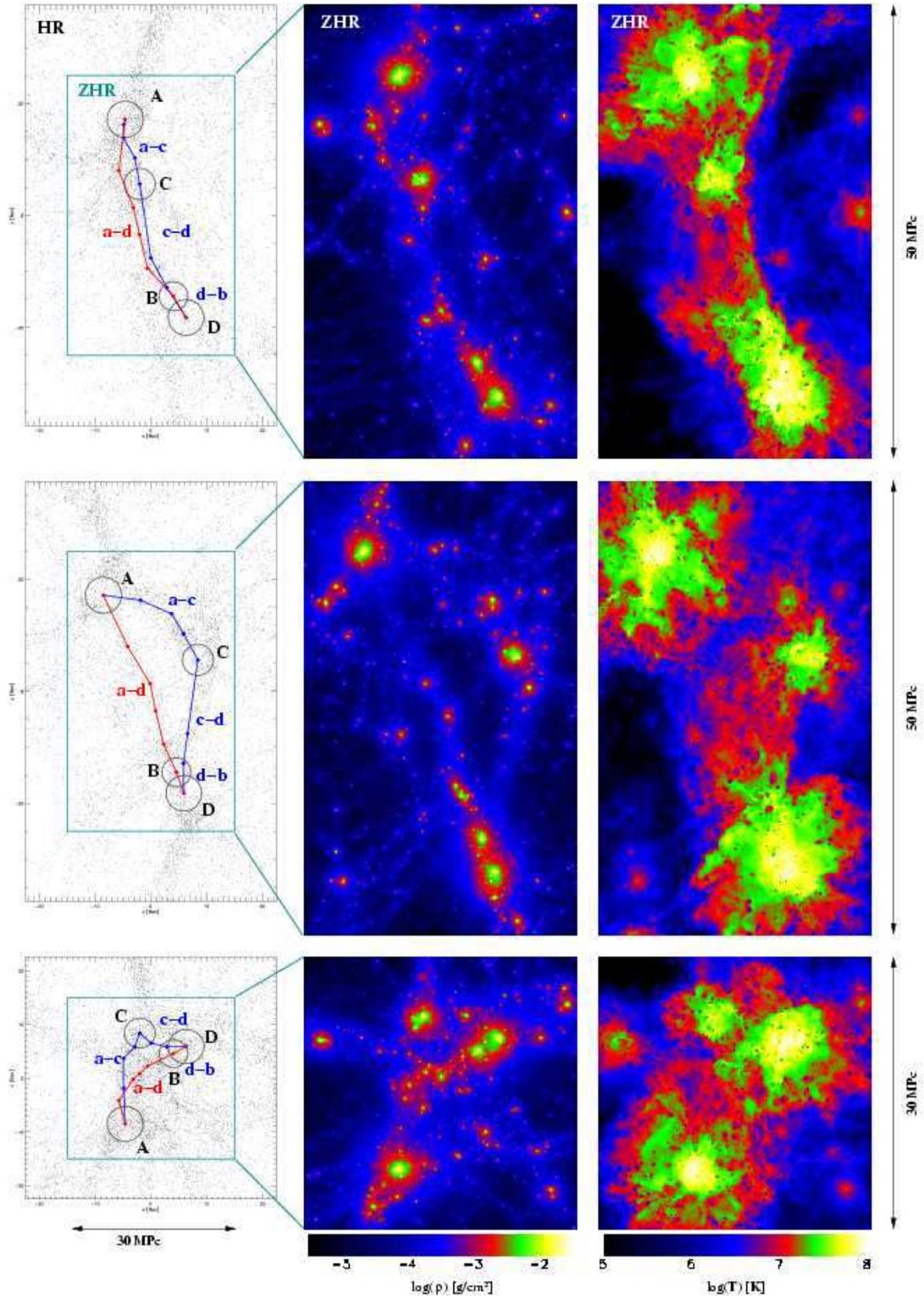}
\end{center}
\caption{Projected maps extracted from the simulation. 
Left column: the distribution of the haloes
containing at least 30 (DM+gas) particles, extracted from a
box of size $50\times 50 \times 70$ Mpc$^3$ placed within the HR
region. Circles indicate the
positions and the virial radii of the four most massive objects
(labeled from $A$ to $D$), while the lines refer to the paths used to
follow the filamentary structures.  The rectangle shows a zoom
into the HR region (ZHR, with size $30\times 30 \times 50$ Mpc$^3$) 
displayed in the other colums, which has no contamination effects
produced by the presence of LR particles.  Central column: the
projected gas density in a logarithmic colour scale (shown at the
bottom).  Right column: the projected mass-weighted temperature, also
in a logarithmic colour scale shown at the bottom.  Each row refers to
a different cartesian projection: from top to bottom the projection is
along the $z$, $x$ and $y$ axes.}
\label{fig:fig1}
\end{figure*}

The distribution of dark matter and gas inside the simulated
supercluster region is quite complex. This is already evident in
Fig.~\ref{fig:fig1}, where we present projected maps obtained from the
results of the final output of the simulation, corresponding to
redshift $z=0$.  Only in the left column we displayed the whole
HR region (i.e.  $50\times 50 \times 70$ Mpc$^3$) to give a better
impression how the supercluster is embedded within its environment. In
the other columns and also in the following analysis we restrict
ourself to a zoom of the high-resolution region (ZHR) of dimensions
$30\times 30 \times 50$ Mpc$^3$. This sub-box contains all the
filaments connecting the four main halos within the supercluster
region. The different rows correspond to the three different cartesian
projections. In the left column, each dot represents a halo
containing at least 30 (DM+gas) particles, i.e.  with a mass larger
than about $5\times 10^{10}\,M_\odot$.  In the same plots the circles
show the positions (and virial radii) of the four most massive galaxy
clusters belonging to the supercluster, which are identified by the
letters {\it A} to {\it D}, starting from the most massive object.
In Table \ref{tab:tab0} we summarize the main properties of these
four clusters computed inside the virial radius, $R_{\rm vir}$
(defined by using the overdensity threshold dictated by the spherical
top-hat model; see e.g. \citealt{eke96}): the virial mass, $M_{\rm
vir}$; the mass-weighted temperature, $T_{\rm MW}$; the X-ray
luminosity computed in the [0.1-10 keV] energy band, ($L_{\rm X}$);
the central value of the Compton-$y$ parameter produced by the SZ
effect (averaged over the three different Cartesian projections). Each
of these four objects is very well resolved containing inside the
virial radius between $1\times10^6$ and $2\times10^6$ particles,
depending on their mass.  The methods applied to compute the last two
`observables' will be briefly described in the following
Sect.~\ref{sect:observations}.  Notice that in this paper we prefer to
use the mass-weighted estimator of the temperature because it is more
related to the energetics involved in the process of structure
formation. As shown in earlier papers
\citep{mathiesen2001,mazzotta2004,gardini2004,rasia2005}, the
application of the emission-weighted temperature, even if it was
originally introduced to extract from hydrodynamical simulations
values directly comparable to the observational spectroscopic
measurements, introduces systematic biases when the structures are
thermally complex, such as in the supercluster region considered here.

\begin{figure*}
\begin{center}
\includegraphics[width=1.0\textwidth]{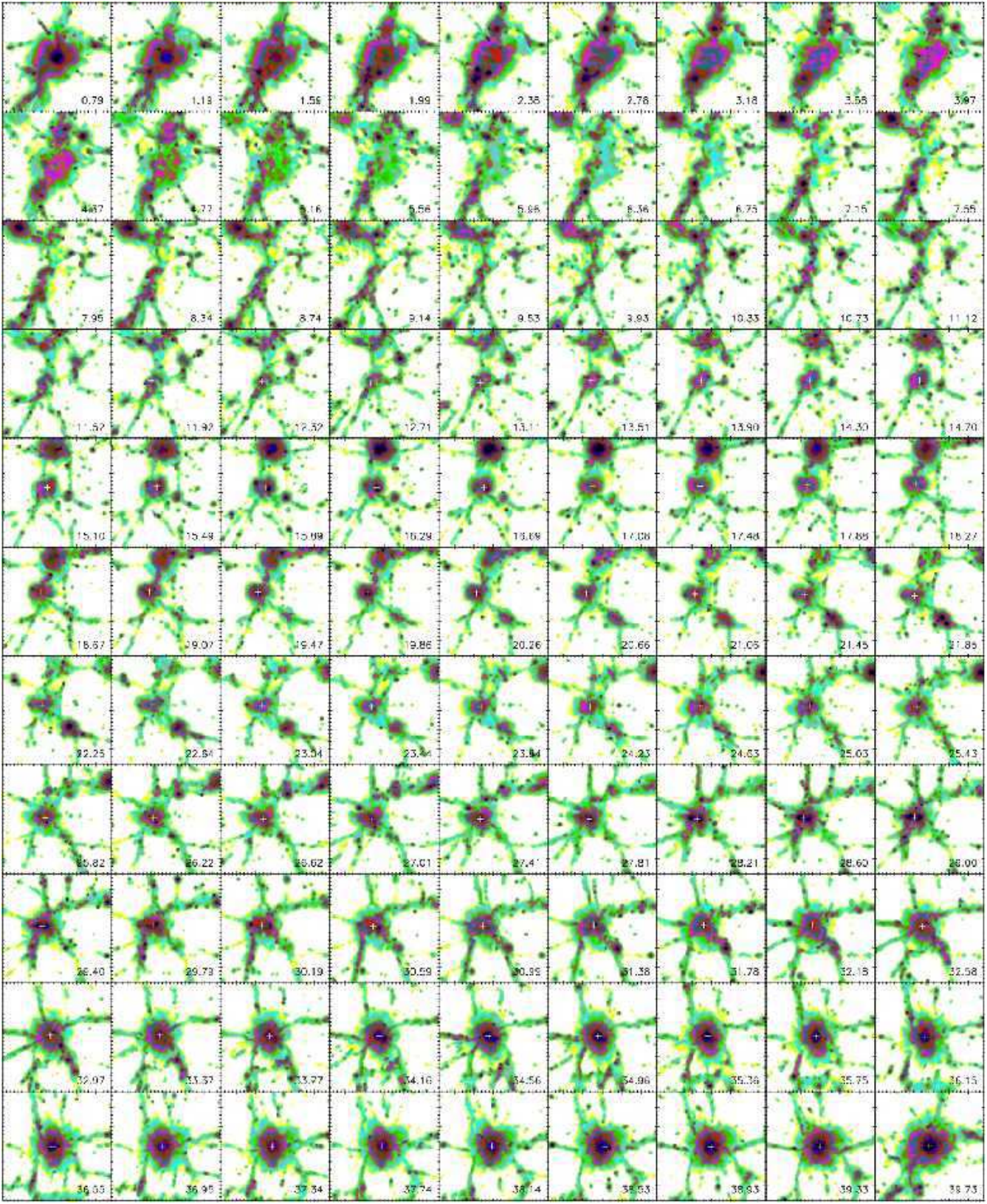}
\end{center}
\caption{
 Contours for gas isodensity levels computed in 100 equidistant slices
 located between haloes $A$ and $B$. The number at the bottom right
 corner indicates the distance (in Mpc) from halo $A$. The white cross
 marks the centre of the filament and it has been shown only for the
 slices for which the symmetry axis of the filament is roughly
 parallel to the line of sight. Notice that this does not occur in the
 slices with distances between $0.79$ and $11.52$ Mpc from cluster
 $A$. The color scale is the same than in the upper panels of Figure
 \ref{fig:fig4a}.}
\label{fig:fig_slices}
\end{figure*}

\begin{table}
  \caption{The main properties of the four largest haloes found inside
  the supercluster region.  Column 1: the halo identification letter.
  Column 2: the virial radius $R_{\rm vir}$ (in units of Mpc).  Column
  3: the virial mass $M_{\rm vir}$ (in units of $10^{15} M_{\odot}$).
  Column 4: the mass-weighted temperature $T_{\rm MW}$ (in units of
  keV).  Column 5: the X-ray luminosity, $L_{\rm X}$, computed in the
  [0.1-10 keV] energy band (in units of $10^{45}$ erg s$^{-1}$).
  Column 6: the central value of the Compton-$y$ parameter, averaged
  over the three cartesian projections (in units of $10^{-4}$).}
  \begin{center}
  \begin{tabular}{c|c|c|c|c|c|}
  \hline
  Halo & $R_{\rm vir}$ & $M_{\rm vir}$ &
  $T_{\rm MW}$ & $L_{\rm X}$ & $y$\\
  identification & (Mpc) & ($10^{15} M_{\odot}$) &
  ($10^8$K) &  ($10^{45}$ erg s$^{-1}$) & ($10^{-4}$) \\
  \hline
   $A$ & 3.25  & 1.89 & 0.69 & 2.08 & 1.82 \\
   $B$ & 3.18  & 1.79 & 0.70 & 1.33 & 1.37 \\
   $C$ & 2.78  & 1.19 & 0.46 & 1.26 & 1.04 \\
   $D$ & 2.58  & 0.95 & 0.55 & 1.42 & 1.59 \\
  \hline
  \end{tabular}
  \label{tab:tab0}
  \end{center}
\end{table}

The panels in the central and right columns of Fig.~\ref{fig:fig1} are
the projected maps for the gas density and mass-weighted temperature,
respectively. In these cases the displayed region (ZHR) corresponds to
the rectangles marked in the panels on the left.  From
the gas distribution, it is easy to recognize the
positions of the four massive clusters, where the projected density
can be larger than $10^{-2}$ g cm$^{-2}$. At the same locations, also
the temperature maps have the highest values, with $T\ge 5\times
10^{7}$ K. In all projections than the thermal structures of the clusters named
$B$ and $D$ appear as overlapping because of their small projected
distance.

\begin{figure*}
\includegraphics[width=1.0\textwidth]{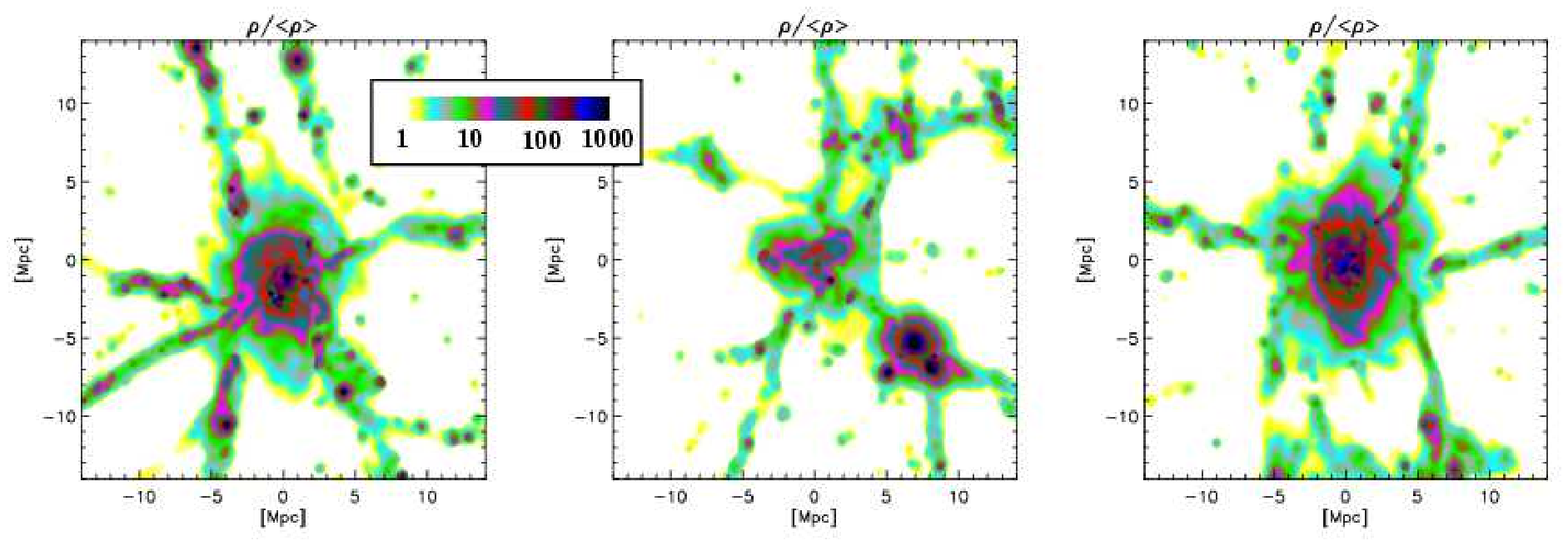}
\includegraphics[width=1.0\textwidth]{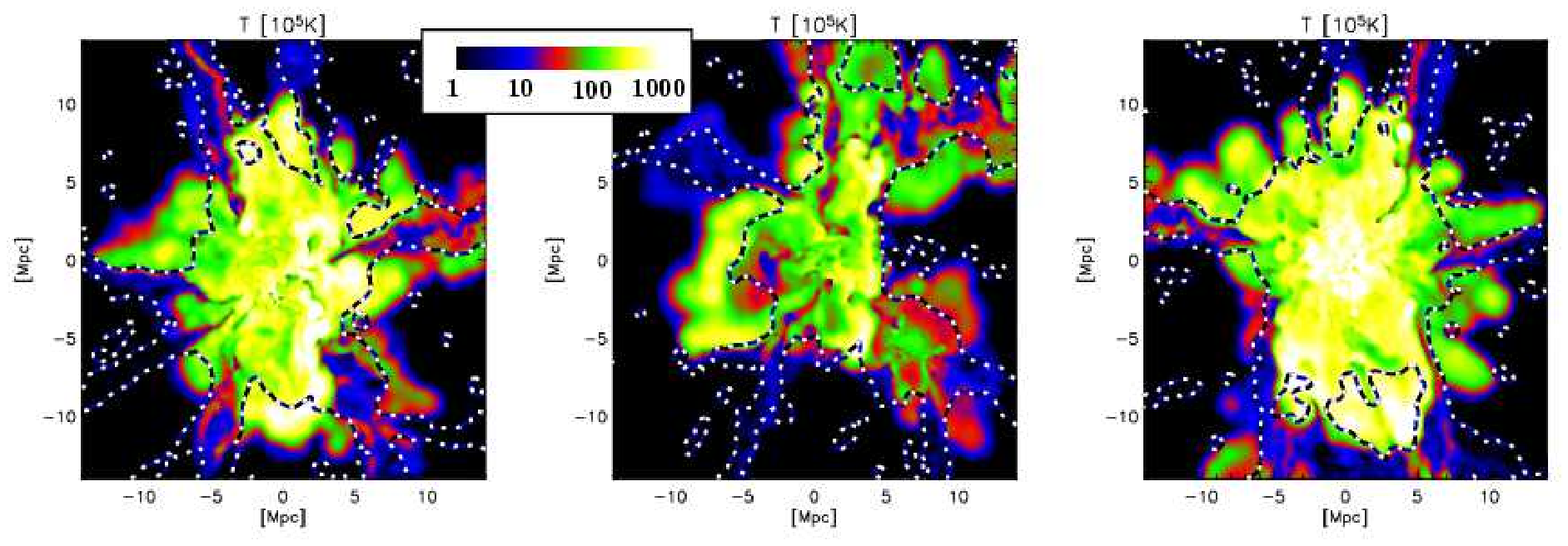}
\caption{
Upper panels: contours for gas isodensity levels computed on a
plane which is located half-way between different cluster pairs and
orthogonal to the bridge direction.  Different panels refer to the
bridges connecting clusters $A$ and $C$ (left), $B$ and $C$ (centre)
and $B$ and $D$ (right).  The logarithmic colour scale (in units of
the mean density) is also reported. The side length of the displayed
regions corresponds to 30 Mpc.  Lower panels: the same but for gas
(mass-weighted) temperature (in units of $10^5$ K).  For reference,
the dashed lines show the gas isodensity contour level corresponding
to unity, already displayed in the upper panels.  }
\label{fig:fig4a}
\end{figure*}

Looking at the distributions of both haloes and gas, the presence of the
so-called cosmic web is evident, i.e. elongated structures (one- or
two-dimensional bridges) connecting the main virialized structures and
spanning the whole region. Along these structures we find that the gas
temperature, the 3D density and the projected density are typically much lower
than $10^{7}$ K, $10^{-28}$ g cm$^{-3}$ and $10^{-3}$ g cm$^{-2}$,
respectively.

As discussed in detail by \cite{colberg05} who analyzed a set of 228
filaments identified in a high-resolution N-body simulation for a
$\Lambda$CDM cosmology, it is not easy to define a shape for this kind
of structures. They found that roughly half of the filaments are
either warped or lie off the line connecting two galaxy clusters.
This is confirmed also in our simulation.  In the left column of
Fig.~\ref{fig:fig1} the paths along which we measure some of the
properties of the filamentary structures, namely the bridge connecting
the clusters $A$, $B$ and $D$, and the bridge connecting the clusters
$A$, $C$ and $B$, are indicated by lines. The paths are found
iteratively. Firstly, we define a straight line connecting the cluster
pairs; then, we cut a slice perpendicular to this line and find the
maximum of the gas density on the slice.  We repeat the same procedure
iteratively, by cutting other slices in between the newly defined
connecting points. The search for new points stops when the positions
of the gas density maxima in two consecutive slices converge within
the resolution limit. The path lines appear to be not perfectly
aligned along the cluster-cluster axis, with the exception of the path
between clusters $B$ and $D$, the closest pair, which has a straight
structure. In general, by analyzing the several bridges present in the
simulation it seems that a typical coherence length for the direction
of these structures is about $\sim 5$ Mpc, but it can reach even
higher values, as shown in the example in Fig.~\ref{fig:fig_slices},
where we present a tomography of the 40 Mpc wide region between haloes
$A$ and $B$, made by using 100 equidistant slices. From the plots it
is clear that the filament departs from halo $A$ in the lower left
corner.  At distances larger than 10 Mpc the elongated structure
starts to bend in the direction of halo $B$.  Then, it stays
approximately parallel to the axis joining the two haloes for more
than 25 Mpc, before it merges with cluster $B$. In different
panels of the tomography it is evident that the filament represents a
junction of several sheets present in the cosmic web. Although the
position of the structures changes along the filament, the tomography
clearly shows that the basic properties of the filament itself are
quite stable for very large distances.  This is particularly evident
along the last 25 Mpc, where the slices of the tomography are roughly
orthogonal with respect to the filament: the extension of the filament
(roughly 5 Mpc in radius), as well as the position and number of
sheets entering it do not vary in a significant way.

The matter bridges present in the projected maps shown in
Figs.~\ref{fig:fig1} and \ref{fig:fig_slices} can correspond both to
two-dimensional (sheets) and one-dimensional (filaments) structures.
These structures have been already studied in previous detailed
analyses of numerical simulations: the results show that in many cases
the filamentary structures are quite complex systems, not always well
defined in one dimension, but often corresponding to projected sheets
or intersections of them
\citep{dave1997,sathyaprakash1998,colberg1999, pierre2000}.  In order
to better understand the true shape of the structures defining the
cosmic web, in the upper panels of Fig.~\ref{fig:fig4a} we show for
three different bridges between clusters the density
distribution of the gas computed on a plane which is orthogonal to the
bridge direction and which is located half-way between the two
clusters defining it. The bridge {\it A-C} (left panel) appears
to be very irregular, with the presence of different matter chains
entering the central structure.  The bridge {\it C-B} (central panel)
can be very well described as the intersection of several
sheets. Finally, the bridge connecting the clusters $B$ and $D$ (right
panel), is more regular and corresponds to a higher gas overdensity
because the two systems are very close and expected to merge in the
near future.

\begin{figure}
\includegraphics[width=0.495\textwidth]{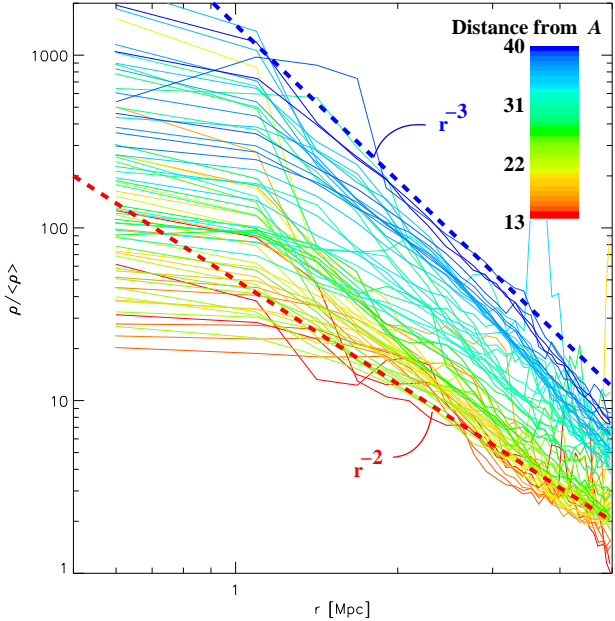}
\caption{
The radial gas density profiles in the filamentary structure
connecting haloes $A$ and $B$ (see Fig.~\ref{fig:fig_slices}), as a
function of the distance from the filament centre. Different coloured
curves refer to slices located at different distances $d$ from cluster
$A$ (see the corresponding colour bar); only results for slices with
$d>13$ Mpc are shown.  The red and blue dashed lines show the
asymptotic trends, $r^{-2}$ and $r^{-3}$, respectively}
\label{fig:fig_profs}
\end{figure}

\begin{figure}
\includegraphics[width=0.495\textwidth]{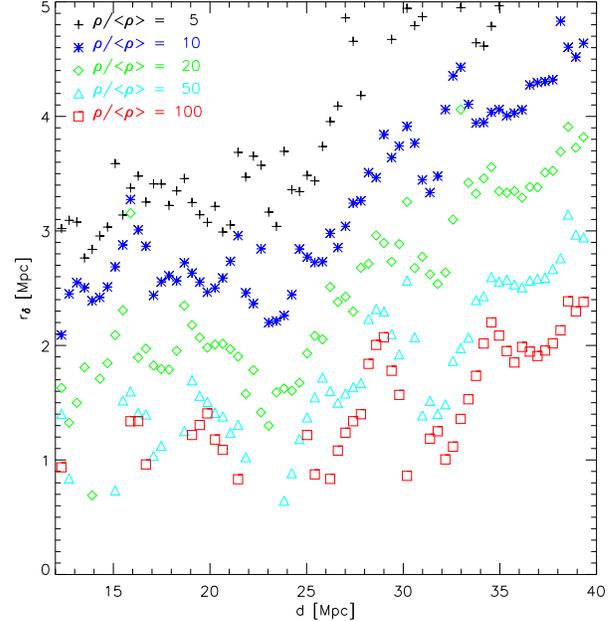}
\caption{The measured thickness $r_\delta$ of the filament shown
in Fig.~\ref{fig:fig_slices}, as function of distance $d$ from halo
$A$. The results are shown for different overdensities, as indicated
in the plot.}
\label{fig:fig_thick}
\end{figure}

Taking advantage of the extension and regularity of the filament
connecting haloes $A$ and $B$, we can measure its radial gas density
profile as a function of the distance from cluster $A$. This
quantity has been plotted in Fig. \ref{fig:fig_profs} for all the
slides shown in Fig.\ref{fig:fig_slices} having a distance larger than
$\approx 13$ Mpc from the cluster A. Each profile is drawn up to 5
Mpc, which represents the typical scale up to which the filament looks
round, and a radial profile can be safely defined: at larger distances
the structure dissolves into sheets of matter, developing the arm-like
structures visible in several slices.  In the plot the density
profiles are coloured according to their position along the filament
(see the colour bar). It is clear that the profiles grow when moving
towards the cluster B, reaching the highest value for the slide
centered on the cluster. In this case the gas profile falls off as
$r^{-3}$ (dashed-thick blue line) in the outer part, as expected from
a Navarro-Frenk-White profile
\citep{NA97.1}: this confirms that the gas behaves as the dark matter
component in the external part.  On the other hand, the profiles computed
along the filament (red and green lines) seem to be less steep: for example,
in the slice located at a distance of about $13$ Mpc from the cluster $A$,
they decrease in the external part like $r^{-2}$ (dashed thick red line).
Also the slopes of the density profiles in the inner part seem to be shallower
along the filament than in proximity to the clusters.  Notice that the
transition between the two regimes occurs at distances which exceed by almost
a factor of four the virial radius of cluster $B$.

The previous result is confirmed by the analysis of the thickness of
the filament, $r_\delta$, defined as the radius at which the profiles
drop below a given density threshold.  In Fig.~\ref{fig:fig_thick} we
show the behaviour of $r_\delta$, as a function of the distance $d$
from $A$. Even if there are some fluctuations produced local
substructures, it is evident that the thickness is roughly constant up
to $d \approx 30$ Mpc, then it monotonically increases when
approaching the cluster $B$.  Notice that this rise starts at a large
distance (larger than 10 Mpc) from the halo.  This result is
independent of the chosen density threshold.

  
\section{Properties of the gas inside the bridges}\label{sect:gasproperties}

\begin{figure*}
\includegraphics[width=0.495\textwidth]{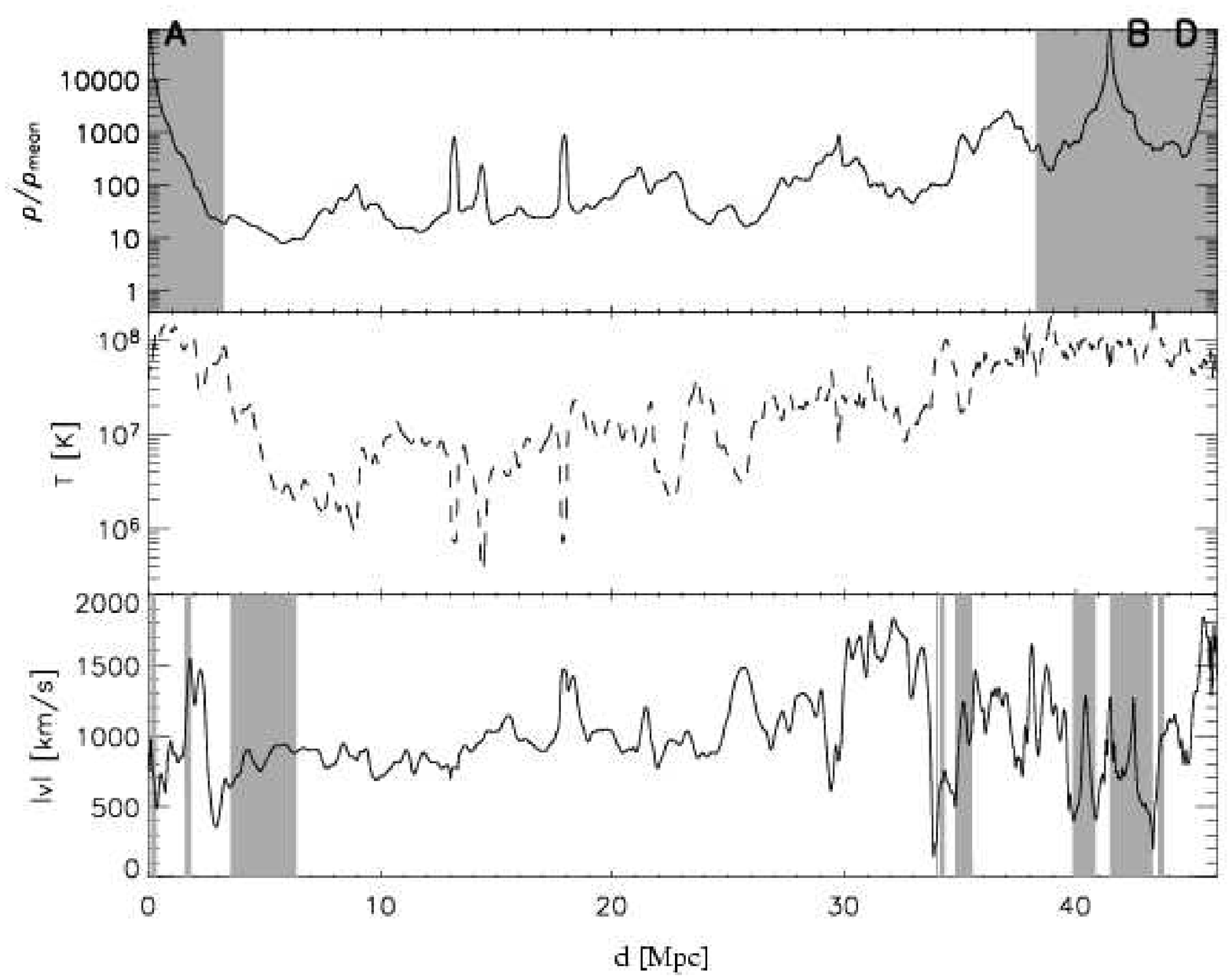}
\includegraphics[width=0.495\textwidth]{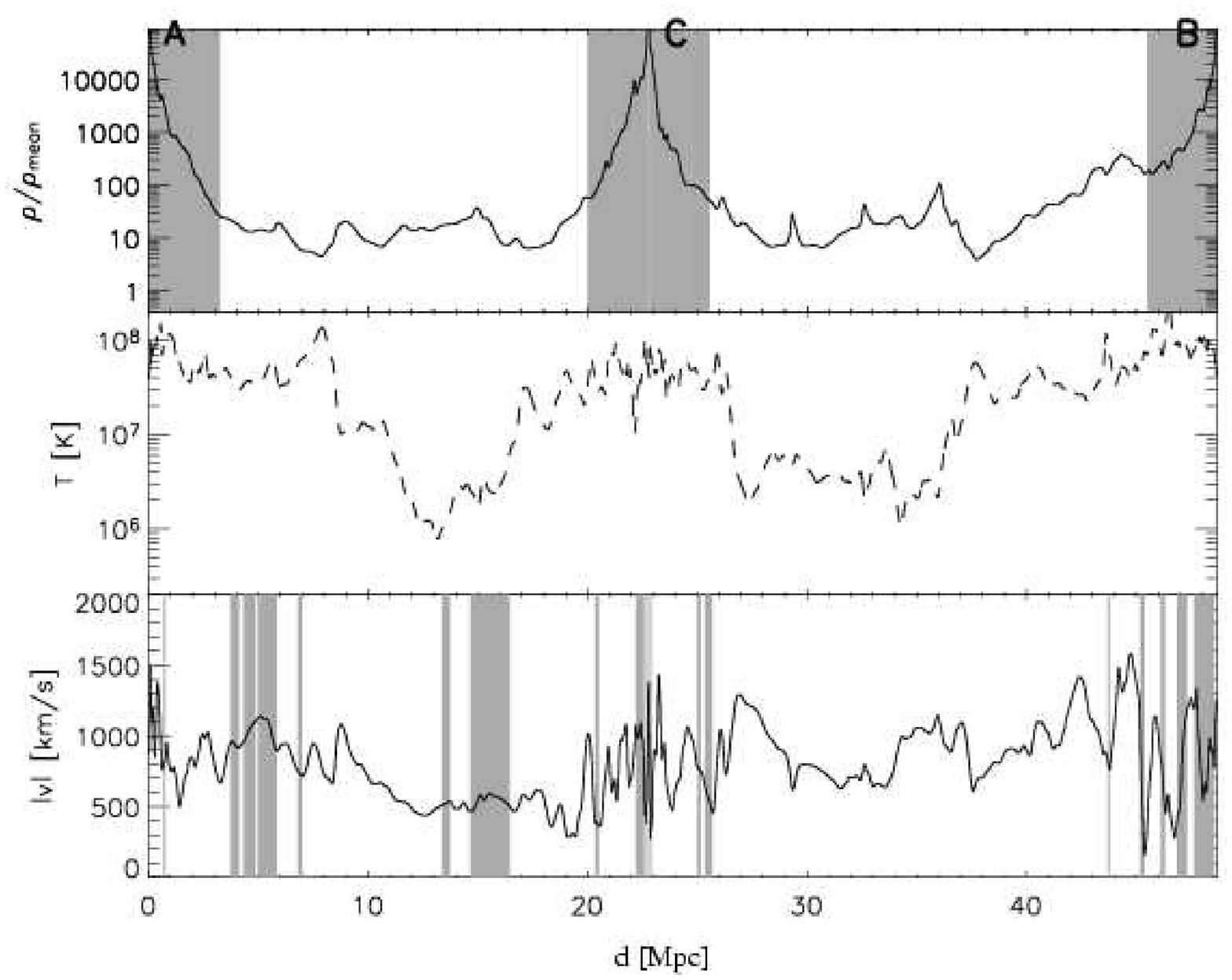}

\caption{The physical state of the gas along the bridges
connecting the galaxy clusters: the paths {\it A-B-D} and {\it A-C-B}
(see for reference the left panels in Fig.~\ref{fig:fig1}) are shown
on the left and right, respectively. Upper panels: the gas density (in
units of the mean density); the grey regions correspond to the virial
radii of different clusters (as labeled by the letters); notice that
clusters {\it B} and {\it D} are so close that their radii appear as
overlapped.  Central panels: the gas (mass-weighted) temperature (in
units of K).  Bottom panels: the modulus of the three-dimensional
velocity field (in units of km/s); the grey regions indicate where the
velocity field is well aligned with the filamentary structure,
i.e. where it forms an angle smaller than 45 degrees with the axis of
the filament.}
\label{fig:fig3a}
\end{figure*}

We now discuss the physical state of the gas contained inside the
bridges.  This is relevant to understand how the filaments and the
sheets can be seen in the optical, X-ray and millimetric bands (see
the next section). We estimate the gas properties making use of the
SPH kernel to interpolate their value at different positions along the
filamentary paths drawn in the left panels of Fig.~\ref{fig:fig1}.
The results for the structures connecting galaxy clusters $A$, $B$ and
$D$ and galaxy clusters $A$, $C$ and $B$ are shown in
Fig.~\ref{fig:fig3a} (left and right panels, respectively).

>From the curves showing the behaviour of the gas density along the
filaments (top panels), the positions of the largest galaxy clusters
are clearly identifiable, as well as their profiles declining towards
their virial radii. In the bridges between them, the gas density stays
relatively constant, with typical values between 10 and 100 times the
mean (baryonic) density. Some spikes are found, which are produced by
smaller structures, where the gas density is larger. The corresponding
masses are of the order of the galaxy masses.

More interesting is the trend shown by the mass-weighted temperature
(central panels). The highest values obviously correspond to the
clusters, with evidence in some cases of a decrement (by a factor 2)
towards the virial radius \citep[see
also][]{bryan1998,thomas2001,borgani2004,rasia2004}.  In other cases
(see the close pair formed by clusters $B$ and $D$ or the regions
surrounding the clusters $A$, $C$ and $B$ in the right panel) the
outward movement of the accretion shocks creates extended, almost
isothermal regions (with $T_{\rm MW}> 5 \times 10^{7}$ K). A
temperature rise corresponding to about two orders of magnitude is
located at distances larger than 2 virial radii. The minimum value in
the bridges is around $10^{6}$ K.  We notice also that in
correspondence of the small structures present in the bridge $A$, $B$,
$D$ (shown on the left) $T_{\rm MW}$ is smaller than in the
surrounding regions: this is due to the fact that these objects formed
before being heated by the accretion shocks of the larger structures.

Finally, we show in the lower panels the velocity field along the
bridges. The displayed quantity is the modulus of the
three-dimensional field, which reaches values larger than 1000 km/s.
The shaded regions indicate where the infall direction of the gas
forms an angle of less than $45$ degrees with the axis of the
filament. The velocity field tends to be aligned to the direction of
the filamentary structure close to the virial radius of the
clusters. This evidences that the gas is flowing towards the clusters
along the bridges. On the contrary, half way between the dominant
clusters the motion tends to be orthogonal to the filaments,
suggesting that they have a net motion with respect to the box.

The complexity of the thermal structure in the bridges is more clearly
evident in lower panels of Fig.~\ref{fig:fig4a}, where we show the
iso-temperature levels as computed on the same planes shown in
the corresponding upper panels.  We find large and sharp gradients in
the temperature structure positioned far from the centre of the
bridge. Again, there is a strong correlation between the features
present in the thermal distribution and those in the gas density. This
is an indication that at $z=0$ the accretion shocks moved out of
bridge, heating also regions at lower density. This process tends to
create an extended smooth isothermal region, as more evident in the
right panel, where the smaller distance between the two dominant
clusters $B$ and $D$ speeds up the mechanism. Since the temperatures
reached in this shock-heating process can be as large as $10^7$ K, it
would be possible that non-negligible effects can be produced on the
observational quantities, as it will be discussed in the next section.

\begin{figure*}
\includegraphics[width=0.8\textwidth]{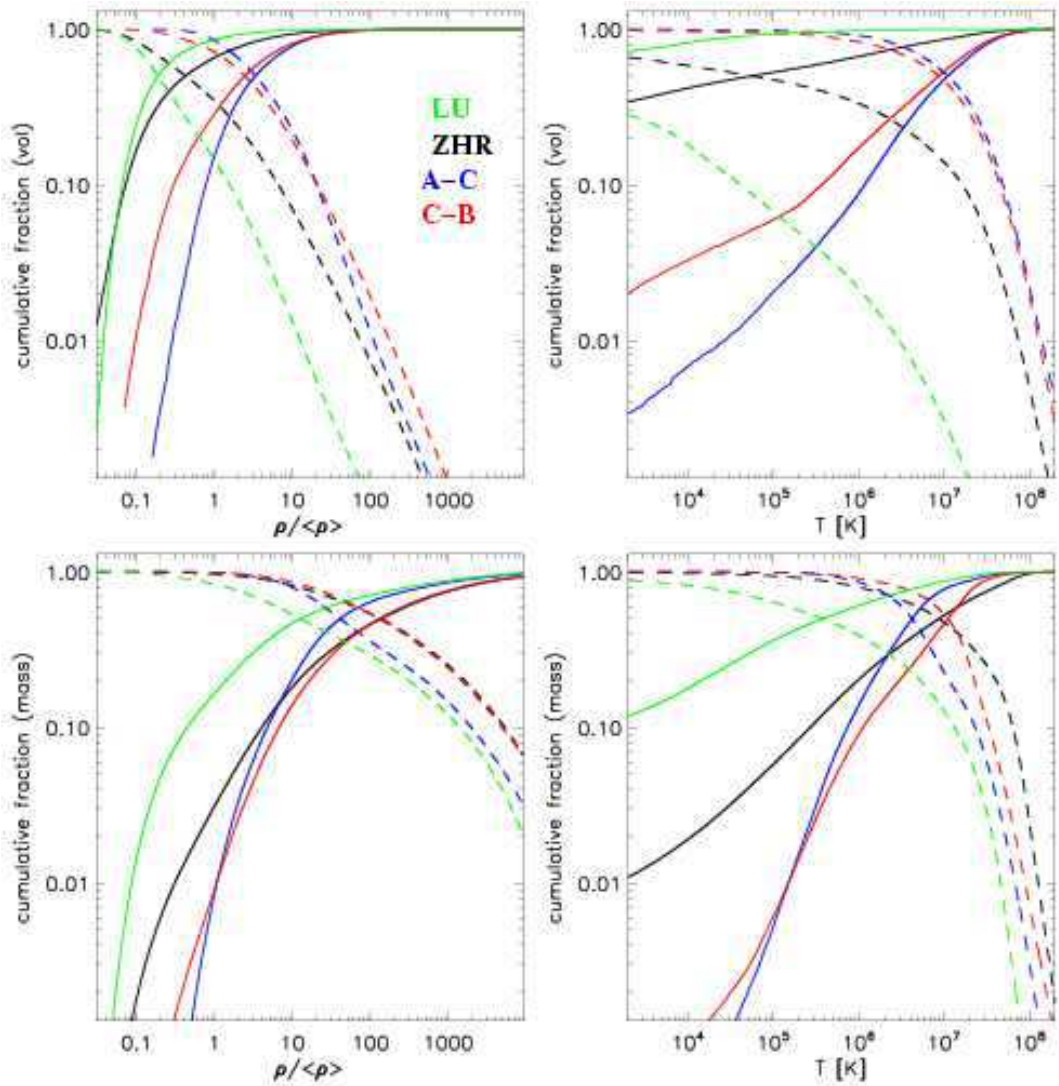}
\caption{Left panels: the increasing solid
(decreasing dashed) curves represent the cumulative fraction of volume
(upper panel) or mass (lower panel) with a gas density smaller
(larger) than a given value.  Different lines refer to different
regions in our simulation: the ZHR box, the bridges A-C and C-B are
shown by black, blue and red lines, respectively.  For comparison we
also show  the results obtained from a different simulation
representing a (70 Mpc)$^3$ box in the Local Universe (LU), centred
at the Milky Way position (green lines).  Right panels: the same as in
the left panels, but for the gas temperature distribution.  }
\label{fig:hist_rho_T}
\end{figure*}

Finally it is quite interesting to quantify the density and
temperature distributions for the gas inside the bridges and compare
them to the corresponding results in more typical regions.  At this
aim we define bridges as cylindrical regions connecting pairs of
galaxy clusters, starting outside $R_{\rm vir}$ and having a radius of
5 Mpc.  In particular we consider the two bridges between clusters A
and C, and between clusters C and B. As examples of different
``typical regions'' we consider (i) the distributions computed in the
whole ZHR region; (ii) a box of 70 Mpc$^3$ extracted from a different
hydrodynamic simulations which was performed using the same numerical
code (assuming the same non-radiative gas-physics and the same
cosmological model). The initial conditions have been constrained in a
way, that the final large scale structure at $z=0$ is representing the
observed Local Universe (LU), with the position of the Milky Way in
the origin. The mass resolution is slightly smaller than that of ZHR
(by a factor of 2.7), but this is not relevant for this comparison.
More details on the LU simulation can be found in
\cite{dolag2005b}. The results for the density and temperature
distributions are presented in Fig. \ref{fig:hist_rho_T} (left and
right panels, respectively). The decreasing (increasing) curves in the
upper panels show the volume fractions having density or temperature
larger (smaller) than a given threshold, while in lower panels we
present the same quantities, but in terms of mass fraction.

Analysing the gas distributions (left panels), we find that the gas
properties in the two bridges A-C and C-B are not very different: for
both 80-90 per cent of the volume and $\approx 99$ of the mass have a
density larger than the mean value, while approximately 20 per cent
 in volume and 80 per cent in mass corresponds to an overdensity of
$\rho/\langle \rho\rangle > 10$.  The different morphology of the
bridges (A-C appears very irregular, while C-B is an intersection of
several sheets; see Fig. \ref{fig:fig4a} and the corresponding
discussion in the previous section) influences the high-density tail:
in fact the bridge C-B displays slightly larger fractions of volume
and mass at overdensity larger than 1000, with respect to the bridge
A-C.  As expected, the two filamentary structures represent regions at
high density when compared to the whole ZHR region, which also includes
extended underdense volumes (about 70 per cent has $\rho/\langle
\rho\rangle < 1$). The differences with respect to the LU simulation
are even larger: in fact it is well know that our Milky Way is
embedded in a slightly underdense region and this is reflected in the
density distributions.  The same trend is found considering the mass
fraction, where however the inclusion of large virialized objects in
the ZHR region and in the LU box enhances the probability of having gas
mass at high density, reducing the differences with respect to the
bridges.

Similar considerations can be done looking to the temperature
distributions (right panels). The thermal structure inside the bridges
is similar, with half of the volume occupied by gas at $T> 10^7$
K. This fraction is much larger than that the corresponding results in
the ZHR region (about 10 per cent) and in the LU simulation (less than
1 per cent). The same trend is evident for the volume-weighted mean
temperature: about $10^7$ K for the bridges, about $10^5$ K in the ZHR
region and less than $10^3$ K in the LU simulation.  This is again an
effect due to the presence in the two last cases of extended
low-density regions, which occupy large volumes and are at low
temperatures. In fact, considering the mass-weighted fractions, the
differences between the bridges and the ZHR and LU regions are reduced:
the mean mass-weighted temperature are approximately $2\times 10^7$,
$10^7$, $5\times 10^6$ and $3\times 10^5$ K, for bridge C-B, ZHR
region, bridge A-C and LU simulation, respectively. In this case, the
difference between the two bridges can be again related to their
different morphology: the sheet intersections for bridge C-B increase
the fraction of gas particles at high density and consequently their
temperature, making it on average denser and hotter than bridge A-C
and ZHR region.

Finally, it is also important to investigate the gas properties in
connection with the WHIM, which is well know to be preferentially
located in the filamentary structures constituting the cosmic web. If
we define the WHIM as the gas having a temperature ranging from $10^5$
and $10^7$ K, we find that it represents 22 and 77 per cent of the
mass contained in the bridges C-B and A-C, respectively.  The large
difference between the two bridges is again produced by their
different morphology.  
From the other side the percentage of mass corresponding to the WHIM
is of order of 50 per cent in both HR region and LU simulation.  In terms of
spatial distribution, the WHIM occupies 37 per cent of the volume in
both bridges, 33 per cent in the whole ZHR region and only 6 per cent
in LU region.


\section{Observables}\label{sect:observations}

In this section we will give an overview over several observables
which are related to the gas and to the dark matter contained in the
bridges.  In particular we will discuss the properties of the halo
distribution, the emission in the X-ray band, the thermal SZ effect
and weak gravitational lensing.

\subsection{Halo distribution}

A first possible way to identify the matter present in the filamentary
structures is to directly investigate the galaxy distribution in
extended redshift surveys.  The detailed study of specific regions of
the universe, like those between close pairs of galaxy clusters, gave
increasing support to the web-like vision of the universe (see, e.g.,
\citealt{pimbblet04a,pimbblet04b,ebeling04}), also suggested by 
wedge diagrams of the large scale structure of the universe obtained
from surveys like SDSS or 2dF.

In our simulation we cannot reproduce exactly this kind of analysis
because in non-radiative hydro-simulations, like ours, it is not
possible to define objects like `galaxies'. In fact, even if the code
follows the evolution of the baryonic component, physical processes
like cooling and star formation are not included.  However, it is
quite safe to assume that the galaxy distribution has to be somehow
related to the distribution of dark matter haloes. For this
reason we investigate the properties of haloes belonging to the
filamentary structure.  In particular we consider the same A-C
and C-B bridges defined in the previous section and we look for their
overdensity with respect to the cosmic mean and the ZHR region.  The
results of our analysis, summarized in Table \ref{tab:tab1}, are
obtained considering objects containing a different minimum number of
member (DM+gas) particles, ranging from 30 to 1000. Adopting the
cosmological baryon fraction, this corresponds to haloes with an
approximate minimum mass ranging from about $3 \times 10^{10}$ and $1
\times 10^{12}$ $M_\odot$.  Notice, however, that especially for small
objects the baryon fraction can vary dramatically, so there is no
longer a direct relation between halo mass and the number of particles
(being the sum of gas and DM particles). We remind that the
bridges were defined using a cylindrical region of a radius of 5
Mpc. This choice does not influence the results we obtained: in fact
the overdensities listed in Table \ref{tab:tab1} change by less than
20 per cent when using a radius of 2.5 Mpc. Our analysis shows
that both bridges are overpopulated by haloes, with typical
overdensities of about 10-14 with respect to the cosmic mean value and
about 5-6 with respect to the whole ZHR region in the simulation.  For
example, we find a number of objects ($\approx 50$) with mass larger
than $10^{12}$ $M_\odot$ in the bridges that is 14 times larger than
what expected on average in the universe.  These values can be
compared with the observed galaxy number overdensity found in
supercluster regions. For example, analyzing a redshift survey of
intracluster galaxies in the central region of the Shapley
concentration supercluster, \cite{bardelli00} find two significant
structures, the first one with an overdensity of about 11 on a scale
of approximately 15 Mpc, the second one with an overdensity of about 3
on a scale of approximately 35 Mpc.

The previous results, together with the projected maps shown in
the left column of Fig. \ref{fig:fig1}, clearly indicate the presence
of a filamentary structure in the halo distribution.  However, the
comparison with real observational data is not completely obvious,
becoming complicated because of two extra difficulties. First, the
observations are using redshifts as distance indicator, which leads to
a systematic distortion due to peculiar velocities.  Second, surveys
are affected by selection effects: only galaxies which are brighter
than a given limiting magnitude are observed. Both problems can
generate a Poisson noise in the attempt of identifying the filamentary
structures.

\begin{figure*}
\begin{center}
\includegraphics[width=1.0\textwidth]{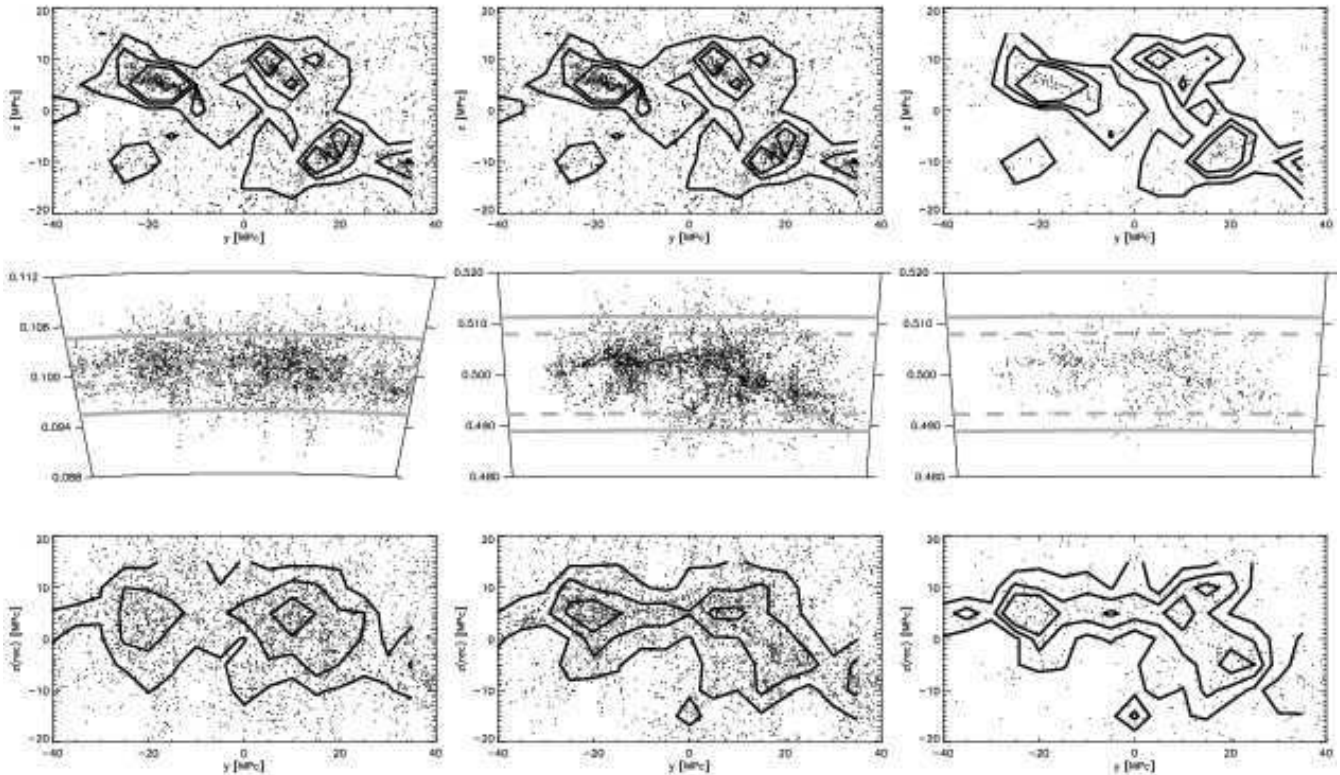}
\end{center}
\caption{The  distribution of `galaxies' in the simulation. 
Panels in the left and central columns refer to a galaxy catalogue
with limiting magnitude $M_{\rm I}=-20$, while panels in the right
column refer to a catalogue with limiting magnitude $M_{\rm I}=-22$.
Upper row: the galaxy distribution in cartesian coordinates, projected
along the $x$-direction. Contour levels show regions with galaxy
overdensities of 1, 2 and 3.  Central row: the corresponding cone
diagram.  The galaxy distributions shown in the upper panels are
displayed in the redshift space, by adding to the real positions the
radial component of the galaxy velocities.  The simulation box has
been placed at a fictitious redshift: $z=0.1$ for the left panel,
$z=0.5$ for the central and right panels.  The solid lines mark the
box size at the assumed redshifts, while the dashed lines correspond
to the region in which haloes not belonging to the box, but having a
peculiar velocity of $v\pm 1000$ km/s could fall: this is uded to
indicate the region of incompleteness due to finiteness of the volume.
Lower row: like the upper panels, but after recovering the cartesian
coordinates from the redshift space.  Again, contour levels show
regions with galaxy overdensities of 1, 2 and 3.}
\label{fig:plothist}
\end{figure*}

In order to investigate these effects we create mock catalogues of
`galaxies' starting from the halo distribution extracted from the
simulation and assuming a one-to-one relation between dark matter
haloes and galaxies.  Following an approach which is standard in
semi-analytic modelling \cite[see, e.g., ][]{delucia2004,croton2006},
we can associate to each object a magnitude by using a
Tully-Fisher-like relation. In particular we adopt the best fit
obtained  by \cite{giovanelli1997} from the analysis of
555 galaxies in 24 clusters:
\begin{equation}
M_{\rm I}-5\log h = -(21 \pm0.02)-(7.68\pm 0.13)(\log W -2.5)\ ,
\end{equation}
where $M_{\rm I}$ is the absolute magnitude in the $I$-band
and $W$ is the velocity
width, which is simply related to the maximum circular velocity
$V_{\rm max}$: $W=2 V_{\rm max}$. 

The distribution of the galaxies present in the HR region is
shown in Fig. \ref{fig:plothist}. In particular the upper panels refer
to the real-space distribution, projected along the $x$-direction, for
two different catalogues, with $M_{\rm I}<-22$ (corresponding to
$V_{\rm max}> 213$ km/s; left and central panels) and $M_{\rm I}<-20$
(corresponding to $V_{\rm max}> 117$ km/s; right panel).
We also show the isodensity contour levels
corresponding to (projected) galaxy overdensities of 1, 2 and 3,
computed by using pixels with size $5\times 5$ Mpc$^2$.  It is easy to
recognize the positions of the highest concentrations, where the four
largest clusters are located, but it is also evident the presence of a
connecting filamentary structure crossing the whole box, in particular
when considering the deepest catalogue.

The previous plots do not include the systematic distortions due to
the galaxy peculiar velocities $v$ which can wash out the filamentary
pattern. To study their effect, we show also the cone diagrams
(i.e. redshift vs. longitude; panels in the central row), produced by
displacing the simulation box at a fictitious redshift and computing
the galaxy positions in the redshift space, i.e. adding to the real
position (converted in recession velocity through the Hubble law) the
radial component of their peculiar velocities.  Since the effect
amplitude depends on the redshift at which we displace the box, we do
two different choices: $z=0.1$ for the panel in the left column,
$z=0.5$ for the panels in the central and right columns.  To point out
regions which can be affected by incompleteness problems related the
finite simulation box, in the plots we also show the box size (solid
lines) and the region in which possible haloes not belonging to this
box, but having a peculiar velocity of $v\pm 1000$ km/s could fall
(dashed lines).  Elongated structures in the observer' direction are
well evident in the left panel showing the galaxy distribution at
$z=0.1$: they are at the angular positions of the main clusters and
correspond to the so-called `fingers-of-God', also observed in real
local surveys. However, due to the limited box size, the cone diagram
is affected by the possible presence of a large number of interlopers.
At higher redshift ($z=0.5$) fingers-of-God are much less prominent,
while the wedge diagrams are dominated by an elongated structure which
extends perpendicularly to the observer' direction, similarly to the
so-called `Great Wall'. The region which is sampled by our simulation
without incompleteness problems is not completely filled by the
galaxies and a filamentary pattern connecting the regions at highest
density can be easily individuated.

Finally, we use the positions in the redshift space to recover the
cartesian distribution of the galaxies. The results are shown in the
lower panels of Fig. \ref{fig:plothist}; again we overlay the
isodensity contours corresponding to galaxy overdensities of 1, 2 and
3.  As expected, the main structures appear significantly washed out,
however the filament can be still clearly identified as an overdensity
larger than 2. Moreover we find that the bright galaxies, although
much less numerous, trace the filament much better.

\begin{table*}
  \begin{center} \caption{ 
The density of haloes found in the bridges. Different columns present
the results as a function of the minimum number of member particles
$N_{\rm min}$, used in the halo definition; $M_{\rm min}$ is the
corresponding approximate minimum mass (in units of $10^{11}
M_\odot$).  Lines 3 and 4 report the mean number density (in units of
Mpc$^{-3}$) in the universe {\protect\citep[adopting the mass function
by][]{sheth1999}} and in the ZHR region, respectively.  Lines 5 and 6
give the overdensity (computed with respect to the cosmic mean value)
found in the bridges $A$-$C$ and $C$-$B$, respectively.  Lines 7 and 8
give the overdensity (computed with respect to the ZHR box) found in
the bridges $A$-$C$ and $C$-$B$, respectively.  }
  
\begin{tabular}{l|c|c|c|c|c|}
  \hline
  $N_{\rm min}$                 & 30    & 50   & 100 & 500 & 1000 \\
  $M_{\rm min}/10^{11} M_\odot$ & 0.28 & 0.46 & 0.93 & 4.65 & 9.30 \\
  \hline
  mean density (Mpc$^{-3}$) &  &  & & &  \\
  in the universe &  0.06 & 0.040 & 0.022 & 0.006 & 0.0032 \\
 in the ZHR region &  0.11 & 0.072 & 0.043 & 0.019 & 0.0076 \\
  \hline
  overdensity w.r.t. cosmic mean & & & & & \\
  bridge $A$-$C$ &  11.0 &  10.4 &  10.4 &  19.0 &    14.0 \\
  bridge $C$-$B$ &   9.7 &   9.5 &   9.8 &  14.9 &    11.2\\
  \hline
  overdensity w.r.t. ZHR region & & & & & \\
  bridge $A$-$C$ &   6.0 &   5.8 &   5.3 &   6.0 &    5.9 \\
  bridge $C$-$B$ &   5.3 &   5.3 &   5.0 &   4.7 &    4.7 \\
  \hline
  \end{tabular}
  \label{tab:tab1}
  \end{center}
\end{table*}

\subsection{X-ray emission}

We consider now the X-ray emission of the gas in the filamentary
structure.  In particular we analyze here two-dimensional maps which
have been created by following a standard procedure which exploits the
SPH kernel to distribute on a pixel grid the desired quantities (X-ray
surface brightness, SZ Compton-$y$ parameter, etc.).  Details on the
method can be found in \cite{roncarelli2006} and \cite{dolag2005}.  

\begin{figure*}
\begin{center}
\includegraphics[width=0.85\textwidth]{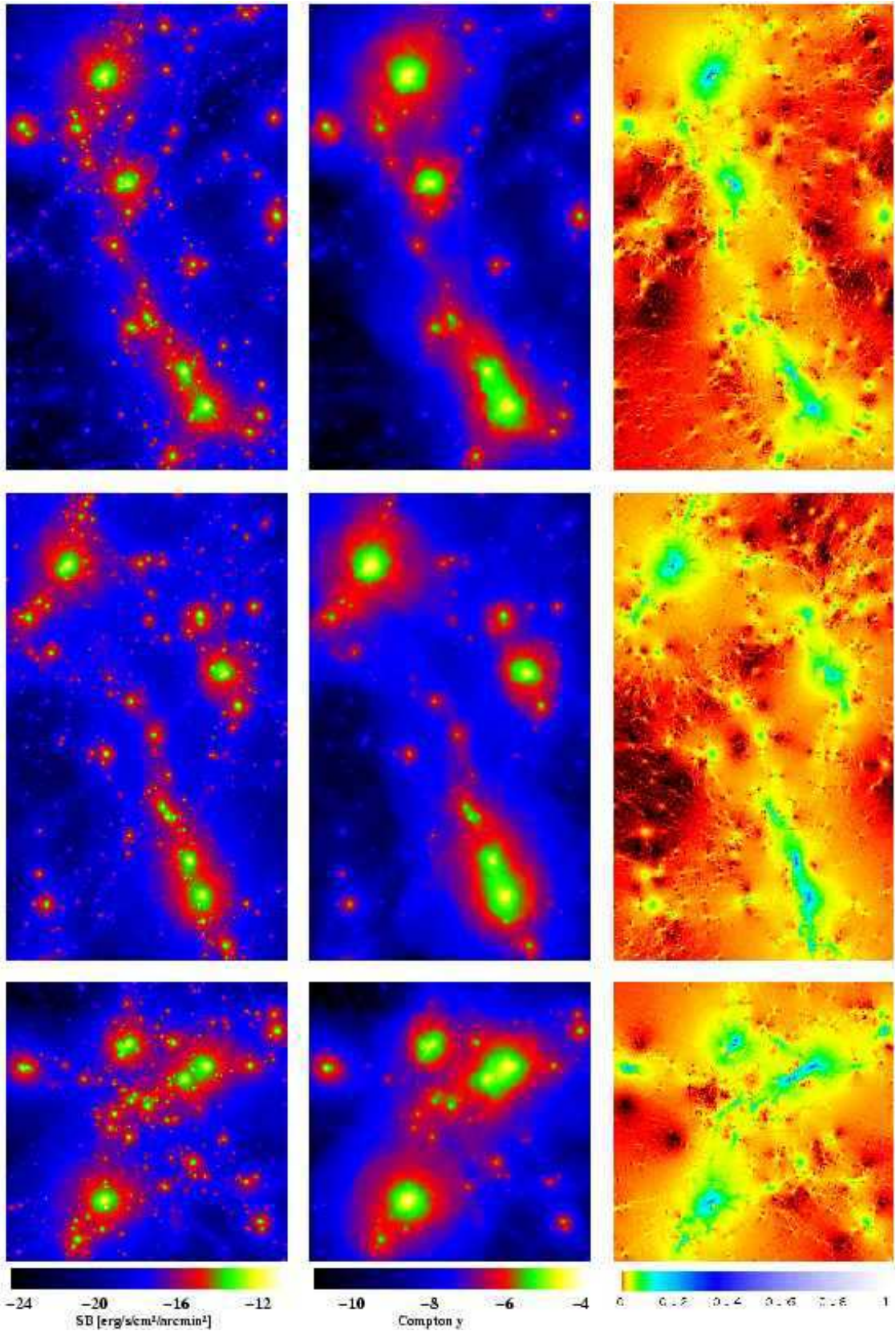}
\end{center}
\caption{Projected maps 
of different observational quantities: the X-ray surface brightness
(SB) produced by thermal bremsstrahlung (left column), the Compton-$y$
parameter quantifying the thermal Sunyaev-Zel'dovich effect (central
column) and the amplitude of the reduced shear produced by weak
gravitational lensing.  The corresponding colour scales (which are
logarithmic in the two first cases) are shown at the bottom.  The
displayed regions are the same as those shown in the rights panels of
Fig.~\ref{fig:fig1}.  Each row refers to a different cartesian
projection: from top to bottom the projection is along the $z$, $x$
and $y$ axes.}
\label{fig:fig2}
\end{figure*}

The X-ray surface brightness (SB) was computed in the
energy band [0.1-10] keV by adopting a cooling function
$\Lambda(T)\propto \sqrt{T}$. Notice that this relation
represents an approximation for temperatures smaller than 2 keV, where
line cooling becomes important.  The resulting maps are shown in the
left column of Fig.~\ref{fig:fig2}.  As expected, the emission is
peaked in correspondence with the largest haloes: the positions of the
four clusters $A$, $B$, $C$ and $D$ are easily recognizable in all
three different projections. These objects appear as embedded in an
extended low-SB region, but with no clear sign of a connecting
structure between them. This is true even for the closest pair,
located in the lower part of the first two panels.  The presence of a
large number of sources, which are not extended and having SB larger
than $10^{-15}$ erg s$^{-1}$ cm$^{-2}$ arcmin$^{-2}$, is also evident
in all maps: they correspond to isolated galaxies and/or small groups.

\begin{figure*}
\includegraphics[width=0.495\textwidth]{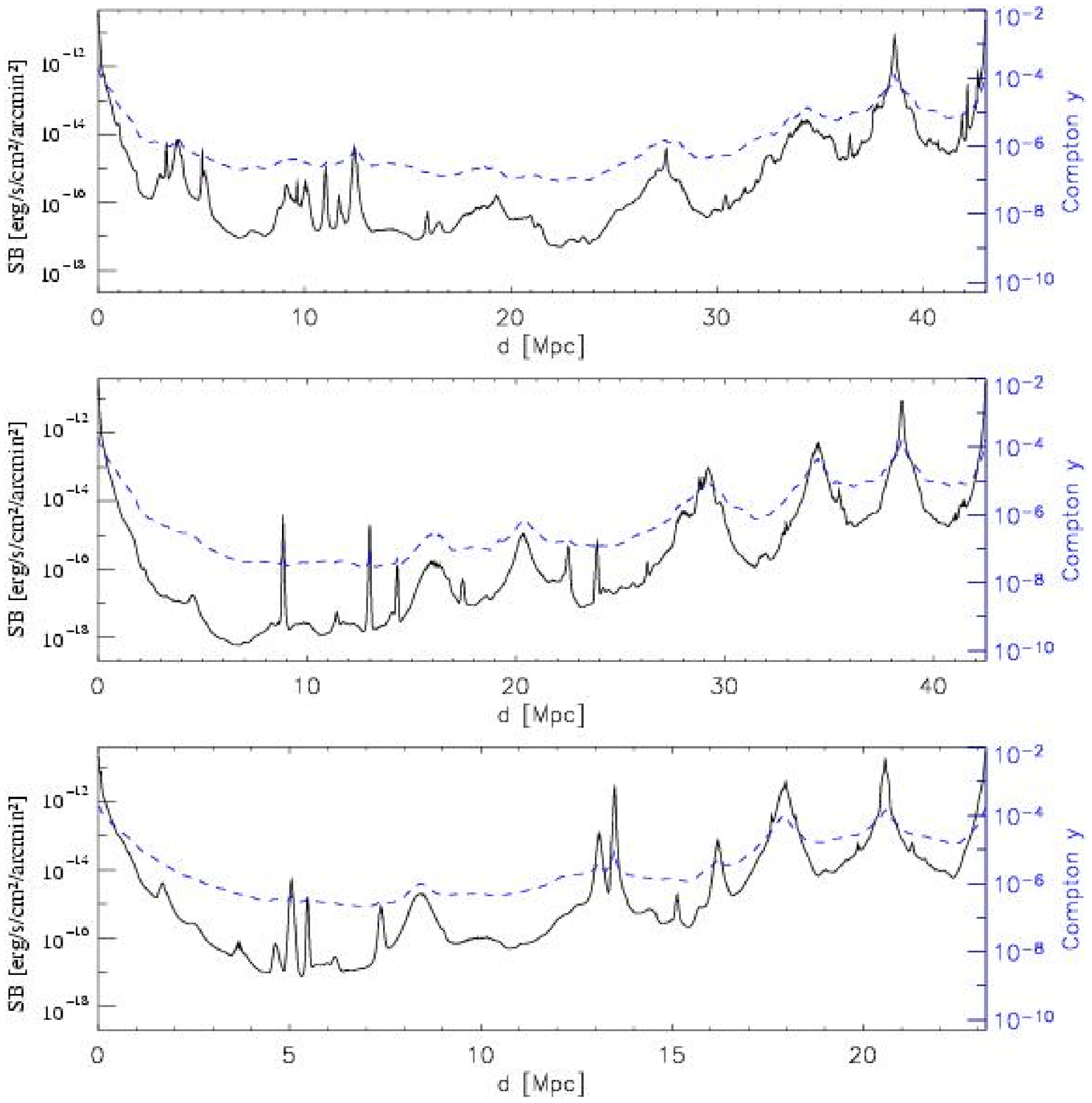}
\includegraphics[width=0.495\textwidth]{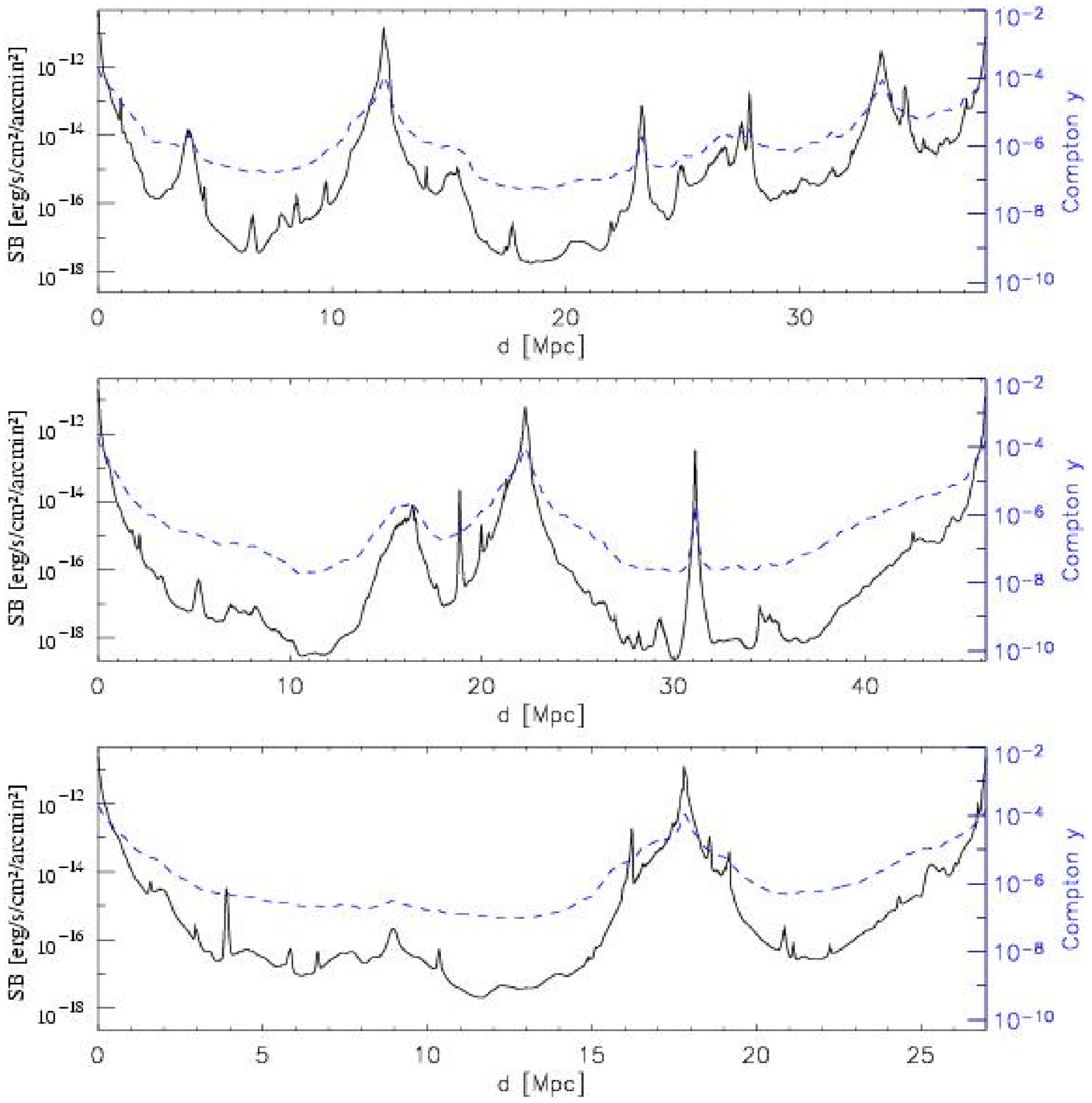}
\caption{The behaviour of the X-ray SB (solid lines with units on the left)
and SZ Compton-$y$ parameter (dashed lines with units on the right)
along the filamentary paths.  Left and right panels refer to the paths
{\it A-B-D} and {\it A-C-B}, respectively.  Each row shows a different
cartesian projection, as in Fig.~\ref{fig:fig2}.}
\label{fig:fig5a}
\end{figure*}

Fig.~\ref{fig:fig5a} shows how the X-ray SB changes along the paths
connecting the clusters, which were defined in the previous sections
(see their positions in Fig.~\ref{fig:fig1}). The results refer to
three different cartesian projections of the supercluster. The SB
along the filaments is at least four orders of magnitude smaller than
in the clusters. Indeed, while the SB corresponding to the largest
haloes is of the order of $10^{-11}-10^{-13}$ erg s$^{-1}$ cm$^{-2}$
arcmin$^{-2}$, its value falls down to $10^{-16}-10^{-18}$ erg
s$^{-1}$ cm$^{-2}$ arcmin$^{-2}$ in the lowest density regions of the
bridges. Nevertheless, as we noticed earlier, many small substructures
produce peaks of X-ray emission, reaching SB of the order of
$10^{-14}$ erg s$^{-1}$ cm$^{-2}$ arcmin$^{-2}$.  Therefore, the
presence of the bridges seems to be more favourably revealed in the
X-rays through compact sources, but certainly not through diffuse
emission. This result is in qualitative agreement with the
analysis made by \cite{pierre2000}, who found that filamentary
structures cannot be clearly detected in the X-ray band because their
signal would be confusion-dominated.

We would remark that our simulation includes non-radiative
hydrodynamics only. It is well known that not considering cooling and
feedback processes tends to change the X-ray emission properties of
diffuse gas, decreasing the brightness in densest regions but
enhancing the emissivity of the low-density material in the filaments
\citep[see, e.g.,][]{bryan2001}. A parallel effect is the production
of groups of galaxies which are overluminous with respect to real data
\citep[see, e.g., the discussion in][]{borgani2004}. This makes the SB
distribution more clumpy, and consequently can bias our estimates of
the X-ray emission.  This excess becomes evident by looking at the
scaling relation between X-ray luminosity $L_{\rm X}$ and temperature
$T$ (not shown): we find that our 27 main haloes scale (with some
dispersion) like $L_{\rm X}
\propto T^2$, as expected for non-radiative simulations, while the
observed X-ray objects follow a steeper relation, in particular at the
mass scales corresponding to galaxy groups
\citep[see, e.g.,][]{osmond2004}.

Even if small, the emission from the filamentary regions could bias,
via projection effects, the correct estimate of the cluster
luminosities.  In order to check this effect we compare the values
obtained by considering the gas particles inside the virial radius
only (see $L_{\rm X}$ reported in Table \ref{tab:tab0}) to the values
obtained by integrating the two-dimensional flux map, which also
include the contribution of particles which appear inside the virial
radius only in projection. In general we find that the increase of the
luminosity is always smaller than 10 per cent. The only exception is
the $y$-projection (last panel), where the two objects $B$ and $D$ are
superimposed and appear as a unique (double) cluster: in this case the
luminosity, which would be assigned to it, is very close to the sum of
the (virial) luminosities of the two components.

\subsection{SZ effect}

In the central column of Fig.~\ref{fig:fig2} we show for the three
different projections the resulting maps for the Compton-$y$
parameter: they are computed by applying a method similar to that
adopted for the X-ray SB, i.e.  using the SPH kernel.  When compared
to the corresponding X-ray maps, the signal appears as coming from
more extended regions: this is the obvious consequence of the
different dependence of the signal amplitude on the gas density.
Unlike in the X-ray maps, the SZ maps reveal a weak evidence of the
existence of two-dimensional structures connecting the most massive
clusters. This is particularly evident in the last projection.  This
general result is confirmed by the trend of the Compton-$y$ parameter
along the paths, shown in Fig.~\ref{fig:fig5a}. The smallest
substructures, from which a significant X-ray emission is coming, are
not revealed by the SZ effect.  On the contrary, the regions
corresponding to the galaxy cluster signal are quite extended, with a
smooth profile which is slowly declining outwards, with no evident
signature of a transition at the virial radius. Along the bridges
connecting the largest objects the Compton-$y$ parameter is very flat,
with values of the order of $10^{-6}$. Outside the filaments, as can
be read off the colour scale in Fig.~\ref{fig:fig2}, the Compton-$y$
parameter drops to values of the order of $10^{-9}-10^{-10}$. 
Notice that similar results have been obtained by \cite{nath2001}, who
estimated the heating of the ICM due to shocks arising from structure
formation, with the aid of the Zel'dovich approximation.

In order to understand the impact of projection effects on the
measurements of the SZ effect by the clusters embedded into the
filamentary structure, we first evaluate $y$ towards the cluster
centres considering only the gas particles which are contained in
their virial regions. The mean of the values obtained considering
three different cartesian projections is reported for each of the most
massive clusters in Table \ref{tab:tab0}. We find that, by varying the
line of sight, the signal changes by up to 20 per cent for cluster $B$
and by up to 7 per cent for the other objects. Recalculating the
Compton-$y$ parameter after including all gas particles within the ZHR
box, we find that further fluctuations of the order of 10 per cent are
possible. In practice changing the line of sight and including the
effect produced by the surrounding filamentary structure can vary the
measurement of the SZ effect for a cluster by up to 30 per cent.

Finally we notice that these estimates of the projection effects
for both X-ray and SZ maps must be considered as lower limits, because
in our analysis we are neglecting possible signals from unrelated
foreground and background objects. This can be quite important for the
SZ effect, where galaxy clusters are more extended structures than in
the X-ray band, and consequently the probability of overlapping is
much higher.

\subsection{Weak gravitational lensing}

Even if the observational detection of filaments via weak
gravitational lensing is still largely uncertain (see the discussion
in the introduction), it is certainly interesting to evaluate its
potentiality by using numerical simulations. These are a useful tool
for calibrating the methods and understand their reliability.  First,
we calculate the deflection angle field around the bridges. The method
used here is described in more detail in a series of previous papers
[see, e.g.,
\citealt{bartelmann98,meneghetti00,meneghetti01,meneghetti03,meneghetti2005}].
We use as lens planes the mass maps obtained by projecting the
three-dimensional mass distribution along the three lines-of-sight
onto a regular grid of $1024\times1024$ cells covering a region of 50
Mpc, adopting the same centre as in the previous analyses. The surface
mass distributions have been smoothed by using the {\em
Triangular-Shaped-Cloud} method \citep{hockney88} for avoiding
discontinuities which might introduce noise in the computation of the
deflection fields.  We propagate through another regular grid on the
lens plane a bundle of $2048\times2048$ light rays and compute for
each of them the deflection angle $\vec{\hat \alpha}(\vec{\theta})$ as
explained in \cite{meneghetti2005}. The reduced deflection angle is
then defined as
\begin{equation}
  \vec \alpha(\vec{\theta})\equiv \frac{D_{\rm ls}}{D_{\rm
  s}}\vec {\hat{\alpha}}(\vec \theta) \ ,
\end{equation}
where $D_{\rm ls}$ and $D_{\rm s}$ are the angular diameter distances
between the lens and the source planes and between the observer and
the source plane, respectively.  For this analysis we shift the
supercluster region to redshift $z_{\rm l}=0.3$. Moreover, we assume
that all sources are at redshift $z_{\rm s}=2$, implying that the
critical surface density,
\begin{equation}
  \Sigma_{\rm crit} \equiv \frac{c^2}{4 \pi G} \frac{D_{\rm s}}{D_{\rm
  l} D_{\rm ls}} = 2.35 \times 10^{15}\, \frac{M_{\odot}}{{\rm
  Mpc}^2}\ ,
\end{equation}
is constant for all of them.  In the former equation, $D_{\rm l}$ is
the angular diameter distance between observer and lens. Although not
realistic, our assumption substantially simplifies the following
analysis and is acceptable because, for a lens at the redshift we
consider here, the critical surface density changes little for typical
sources having redshift $z > 1$. For example, it varies by less than
$25$ per cent when the source plane is moved from $z_{\rm s}=1$ to
$z_{\rm s}=3$.
  
>From the deflection angles $\vec \alpha(\vec{\theta})$, we can easily
calculate both the convergence,
\begin{equation}
  \kappa(\vec \theta)=\frac{1}{2} \left( \frac{\partial \alpha_{1}} {\partial
  \theta_{1}}+\frac{\partial \alpha_{2}}{\partial \theta_2} \right)  
\end{equation}
and the two components of the shear,
\begin{eqnarray}
  \gamma_1(\vec{\theta}) &=& \frac{1}{2}\left(\frac{\partial
  \alpha_{1}}{\partial 
  \theta_1}-\frac{\partial \alpha_{2}}{\partial \theta_2}\right) 
        \label{equation:shear1} \\
  \gamma_2(\vec{\theta}) &=& -\frac{\partial \alpha_{1}}{\partial
  \theta_2}= -\frac{\partial 
  \alpha_{2}}{\partial \theta_1} \ .
        \label{equation:shear2}
\end{eqnarray}
We can then construct the complex shear as
\begin{equation}
  \gamma=\gamma_1+i\gamma_2 \ .
\end{equation}
Finally the complex reduced shear is given by
\begin{equation}
  g(\vec \theta)=\frac{\gamma(\vec \theta)}{1-\kappa(\vec \theta)} \ .
\end{equation}
This quantity is the expectation value of the observed ellipticity
$\chi$ of the galaxies weakly distorted by the lensing
effect. Measuring $\chi$ we thus obtain an estimate of the reduced
shear, provided that we are in the weak lensing regime. The maps of
the amplitude of the reduced shear obtained with this method and
referring to the same regions investigated before are shown in the
right panels of Fig.~\ref{fig:fig2}.  Along the bridges, the reduced
shear grows towards the clusters: in the cluster outskirts, it ranges
between $\sim 0.05$ and $\sim 0.2$, while along the filaments it
decreases to values of $\sim 0.01-0.02$.

\begin{figure*}
\begin{center}
\includegraphics[width=0.58\textwidth]{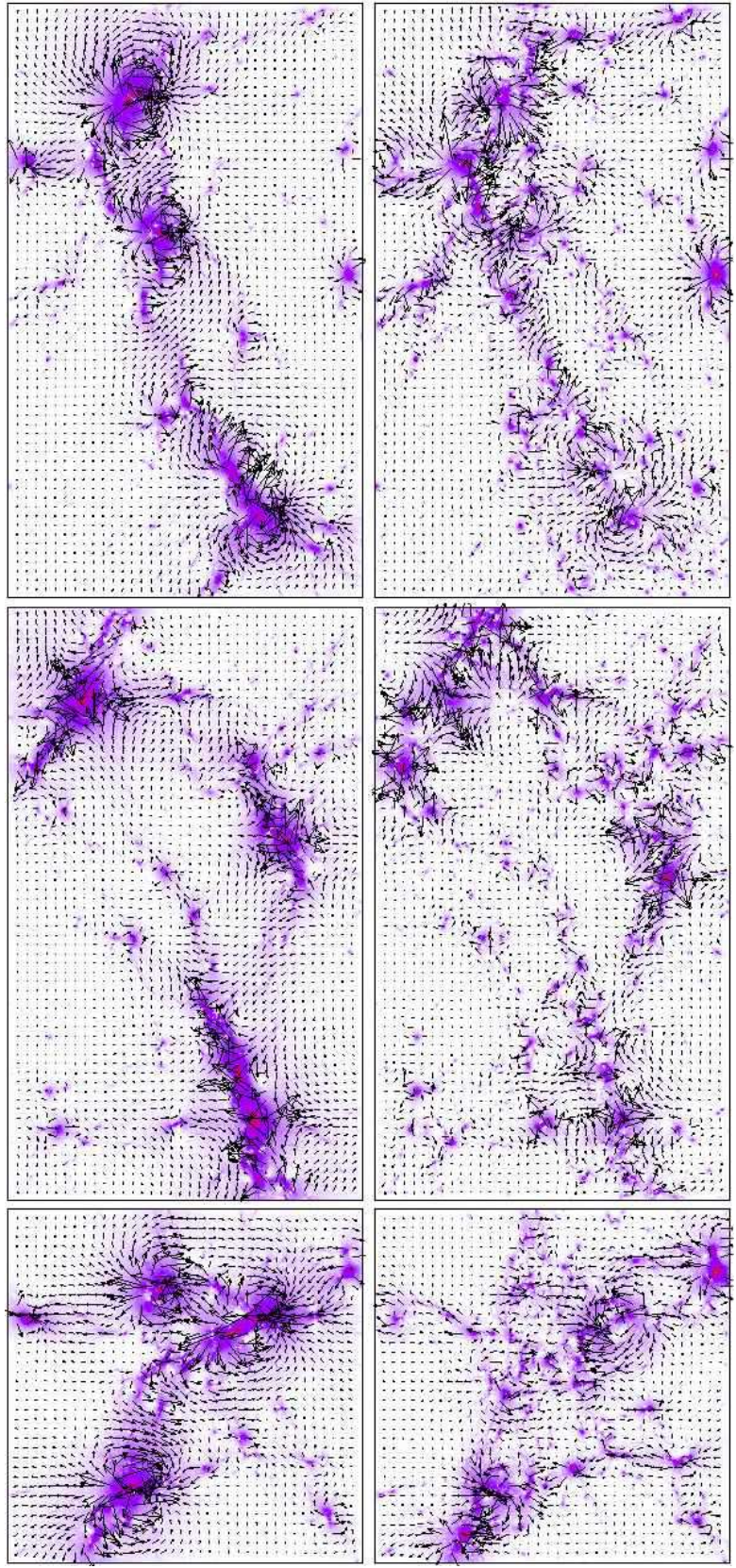}
\end{center}
\caption{
Maps for the reduced shear.  Vectors (in arbitrary units) show the
shear directions and are superimposed on the map of the absolute
amplitude of the signal (already presented in the left column of
Fig.~\ref{fig:fig2}). The displayed regions are the same presented in
the right panels of Fig.~\ref{fig:fig1}.  Each row refers to a
different cartesian projection: from top to bottom the projection is
along the $z$, $x$ and $y$ axes.  The left column shows the total
signal, while the right column the results obtained after removing the
27 most massive haloes identified in the simulation.  }
\label{fig:fig3}
\end{figure*}

Since the shear from the galaxy clusters is so significantly dominant
with respect to the shear caused by the filamentary structure, the
distortion of the background galaxies is mainly tangential to the
cluster concentrations rather than to the bridges of matter connecting
them. This is evident in the left panels of Fig.~\ref{fig:fig3}, where
the shear vectors are superimposed on the shear maps, previously
shown.  The same kind of distortion could in principle be produced by
isolated clusters, i.e. not connected by any filament of
matter. Together with the very low amplitude of the signal, this
represents the more serious obstacle to the detection of the cosmic
web via lensing effects.

In order to better quantify the filament signal, we repeat the same
analysis but considering now what remains if the clusters in our field
are removed, i.e. if the shear produced by their mass distributions is
subtracted from the total shear.  This is done by identifying in the
simulation the particles within the virial radii of the individual
clusters and using them for producing the deflection angle maps to be
subtracted from the previous ones.  Dealing with observations, we
might try to provide a model for each individual cluster, calculate
the shear it produces and remove it from the observed distortion of
the background galaxies. In practice, as shown by \cite{dietrich2005},
it is very complex to find a way for separating the filaments from the
clusters they connect, especially when considering close pairs of mass
clumps.  The results of this analysis are shown in the right panels of
Fig.\ref{fig:fig3}. Here, we removed the shear produced by the
particles belonging to the 27 most massive haloes found in the
simulation, leaving the contributions from the filamentary structures
and from the smallest clumps of matter only. The length of the vectors
showing the shear orientation as well as the colour scale of the
underlying shear maps are arbitrary. Along part of them the shear now
turns to be tangential to the bridges, but still it is locally
tangential to the remaining small mass concentrations.

On the basis of these simple considerations and considering the fact
that the amplitude of the shear produced by the bridges is comparable
to that provided by the large-scale structures along the line of sight
to the distant galaxies, we can conclude that the detection of
filamentary structures through their weak lensing signal is rather
improbable. \cite{dietrich2005} have recently proposed that techniques
based on aperture multiple moments might be used for detecting
filaments between close pairs of clusters. However, we believe that
even this method seems to be difficult to apply for two main reasons:
first, the noise produced by even the smallest clumps distributed or
projected along the filament is large compared to the strength of the
lensing signal from the filament itself; second, contaminations from
the large-scale structure along the line of sight cannot be removed by
using these techniques.


\section{Conclusions}\label{sect:conclusions}

In this paper we discussed the physical and observational properties
of the gas located inside the network of filaments constituting the
cosmic web.  For this aim, we used of the TreeSPH code {\small
GADGET-2} \citep{springel2005} to carry out a high-resolution
hydrodynamical re-simulation of a supercluster-like
structure. Suitable initial conditions were created on a specific
region measuring about $50\times50\times70$ Mpc$^3$, which was
selected from a parent DM-only cosmological simulation because it
contains several massive clusters connected by bridges and sheets of
matter.  The new simulation followed the evolution of the cosmic
structure within the high-resolution region, reproducing the
environmental properties of the larger cosmological box, from where it
was taken. For the first time, it was possible to study at high
resolution the formation of filaments between very massive clusters,
taking non-radiative gas physics into account. The region selected
contains 27 haloes with mass larger than $0.7\times 10^{14} M_\odot$,
four of which with $M_{\rm vir}\magcir 10^{15} M_\odot$.  The
characteristics of the low-density filaments connecting the largest
haloes were analyzed in more detail, with the aim of understanding how
these filamentary structures may appear in view of their galaxy
spatial distributions, X-ray emission and weak lensing signal. Indeed,
these are properties which may be used to probe the existence of the
cosmic web.

In particular we found:
\begin{itemize}
\item the cosmic web is formed by both two- and one-dimensional structures
  (sheets and filaments) which have a quite complex shape; the
  structure of the filaments along their length is characterized by
  changes in both the orientations of the axis of symmetry and in the
  thickness.  The coherence length seems to be typically of the order
  of $\sim 5$ Mpc, but some segments of filaments can extend also to
  $\sim 25$ Mpc without significant changes in direction;
\item by slicing a filament along its axis of symmetry, we find that these
  structures have diameters of the order of $\approx 3-5$ Mpc, within
  which the density contours appear round. Outside of this radius
  sheets of matter depart from the main structure. The thickness of
  the filament grows while approaching a cluster. The growth begins
  already at distances corresponding to $3-4$ times the virial radius
  of the cluster;
\item the radial density profiles along 
  filaments falls off less steeply than in galaxy clusters. The
  profile tends to become less steep, while the distance from the
  clusters increases. At large radii it gradually changes from being
  $\propto r^{-3}$ to being $\propto r^{-2}$;
\item
  the typical values of the gas density in the bridges are between 10
  and 100 times the cosmic mean value, increasing towards the clusters
  they bind and being smaller far from them;
\item
  the thermal structure of the cosmic web is also quite complex,
  depending on the position of the accretion shocks, which, thanks to
  their movement outwards the galaxy clusters, tend to produce
  extended isothermal regions with $T_{MW}> 5 \times 10^7$ K extending
  to the filamentary structures;
\item
  the velocity field also reflects the dynamics of structure
  formation, being orthogonal (aligned) to the filaments at large
  (small) distances from galaxy clusters.
\end{itemize}

In order to discuss the observable properties of the matter and the
gas in the cosmic web, we produced maps of the halo distribution
(which approximately reproduce the spatial distribution of galaxies),
of the X-ray emission, of the thermal SZ effect and of the signal
produced by the weak lensing effect. The main results of their
analysis can be summarized as follows:

\begin{itemize}
\item 
  inside the cosmic web the number density of haloes is about 
  10-14 times larger than the cosmic mean and 4-6 times larger than
  in the neighbourhood, almost irrespectively of the limiting mass
  considered. Galaxies seem to trace quite well the shape of the
  bridges between clusters, even when selection effects and
  redshift-space distortions are considered;
\item 
  in the X-ray maps the emission is peaked on galaxy clusters, which
  are surrounded by extended low-flux regions; along the filamentary
  structures, the surface brightness reaches at most $10^{-16}$ erg
  s$^{-1}$ cm$^{-2}$ arcmin$^{-2}$, except in the positions of small
  haloes (corresponding to isolated galaxies or groups), where the
  surface brightness can be even orders of magnitude larger. In the
  X-rays, the filaments thus appear as superpositions of relatively
  compact sources rather than as an extended region of diffuse
  emission;
\item
  in the maps for the Compton-$y$ parameter, we found the evidence of
  a signal (with values of about $10^{-6}$) coming from the
  two-dimensional structures, connecting the main clusters. The signal
  comes from a more extended region compared to the X-ray emission,
  due to the weaker dependence of the Compton-$y$ parameter on the
  density;
\item 
  we estimated the typical bias on some observational properties of
  galaxy clusters produced by the projection of the surrounding
  filamentary network: we found an increase of less than 10 per cent
  for the X-ray luminosity and up to 30 per cent for the central
  Compton-$y$ parameter;
\item
  we computed the reduced shear in the maps of the weak lensing
  signal, finding along the filaments very low values ($\sim
  0.01-0.02$).  The shear is rarely tangential to the filamentary
  structure.  Rather, it is very often tangential to mass
  concentrations along the filamentary structure. Therefore, it seems
  very difficult to reveal the presence of filaments through weak
  lensing, even those connecting relatively close clusters.
\end{itemize}

In conclusion, our analysis confirms that filaments are very complex
structures. They appear with very different features when looking at
their galaxies, their gas or the distribution of their dark matter. In
this paper we explored several observables which might be used to
reveal the presence of the cosmic web. However the results here
presented, in particular the low fluxes in the X-ray, the weak SZ
effect and the very weak tangential shear signal, suggest that a
conclusive detection of the filamentary structure will be a
challenging problem also for future observations.


\section*{acknowledgements}
Computations were performed on the IBM-SP4 at CINECA, Bologna, with
CPU time assigned under an INAF-CINECA grant.  KD acknowledges partial
support by a Marie Curie Fellowship of the European Community program
``Human Potential'' under contract number MCFI-2001-01227. ER thanks
the MPA for hospitality. We thank the anonymous referee for 
constructive  comments.  We are grateful to
M. Bartelmann, S. Borgani, B. Ciardi, S. Ettori, P. Mazzotta,
Y. Mellier, A. Morandi and M. Roncarelli for useful discussions.

\bibliographystyle{mn2e} 

\bibliography{master2}

\label{lastpage}
\end{document}